\documentclass[preprint,aps,amsmath,nofootinbib,tightenlines]{revtex4-1}

\pdfoutput=1
\usepackage{graphicx}
\usepackage{bm}
\usepackage{graphics}
\usepackage{amsfonts,amssymb}
\usepackage{color}
\usepackage{hyperref}

\def\Dsl{\hbox{/\kern-.6000em D}} 

\def\dsl{\,\raise.15ex\hbox{/}\mkern-13.5mu D}

\def\lqcd{\Lambda_{\rm QCD}}

\def\ltap{\ \raise.3ex\hbox{$<$\kern-.75em\lower1ex\hbox{$\sim$}}\ }
\def\gtap{\ \raise.3ex\hbox{$>$\kern-.75em\lower1ex\hbox{$\sim$}}\ }
\def\OMIT#1{}

\def\lsim{\mathrel{\raise.3ex\hbox{$<$\kern-.75em\lower1ex\hbox{$\sim$}}}}
\def\gsim{\mathrel{\raise.3ex\hbox{$>$\kern-.75em\lower1ex\hbox{$\sim$}}}}

\def\msb{{\overline{\rm MS}}}

\newcommand{\nn}{\nonumber}

\newcommand{\bmS}{\mathbf S}

\newcommand{\ord}{{\cal O}}
\newcommand{\as}{\alpha_S}
\newcommand{\au}{\alpha_U}
\newcommand{\ah}{\alpha_h}

\def\msb{{\overline{\rm MS}}}

\catcode`\@=11
\def\slash{\mathpalette\make@slash}
\def\make@slash#1#2{\setbox\z@\hbox{$#1#2$}%
  \hbox to 0pt{\hss$#1/$\hss\kern-\wd0}\box0}
\catcode`\@=12 

\newcommand{\be}{\begin{equation}}
\newcommand{\ee}{\end{equation}}
\newcommand{\bea}{\begin{eqnarray}}
\newcommand{\eea}{\end{eqnarray}}

\newcommand{\rot}{\color{red}}
\newcommand{\blau}{\color{blue}}

\definecolor{orange}{rgb}{1,0.5,0}
\definecolor{lila}{rgb}{0.5,0,0.5}
\definecolor{brown}{rgb}{0.6,0.4,0.2}
\newcommand{\orange}{\color{orange}}
\newcommand{\lila}{\color{lila}}

\begin{document}


\preprint{ \vbox{ 
\hbox{UWThPh-2012-19}
\hbox{IFIC/12-62}
\hbox{DESY 12-155}
}}

\title{\phantom{x}\vspace{0.5cm}
Renormalization Group Improved 
Bottom Mass \\
from $\Upsilon$ Sum Rules at NNLL Order\\
\vspace{1.0cm} }

\author{Andr\'e~H.~Hoang\footnote{Electronic address: Andre.Hoang@univie.ac.at}$^1$, Pedro~Ruiz-Femen\'ia\footnote{Electronic address: Pedro.Ruiz@ific.uv.es}$^{1,2}$ and Maximilian~Stahlhofen\footnote{Electronic address: maximilian.stahlhofen@desy.de}$^{1,3}$}
\affiliation{
$^1$University of Vienna, Faculty of Physics, Boltzmanngasse 5, A-1090 Wien, Austria \vspace{0.5cm}\\
$^2$IFIC, Universitat de Val\`encia - CSIC, Apt. Correus 22085, E-46071 Valencia, Spain \vspace{0.5cm}\\
$^3$DESY Theory Group, Notkestra\ss e 85, D-22607 Hamburg, Germany
\vspace{1.5 cm}
}


\begin{abstract}
\vspace{0.5cm}
\setlength\baselineskip{18pt}

We determine the bottom quark mass from non-relativistic large-$n$ $\Upsilon$ sum
rules with renormalization group improvement at next-to-next-to-leading
logarithmic order. We compute the theoretical moments within the vNRQCD
formalism and account for the summation of powers of the Coulomb singularities
as well as of logarithmic terms proportional to powers of $\alpha_s\ln(n)$. 
The renormalization group improvement leads to a substantial stabilization of
the theoretical moments compared to previous fixed-order analyses, which did not
account for the systematic treatment of the logarithmic $\alpha_s\ln(n)$ terms,
and allows for reliable single moment fits. For the current world average of
the strong coupling ($\alpha_s(M_Z) = 0.1183 \pm 0.0010$) we obtain 
$
M_b^{1S}=4.755\pm 0.057_{\rm pert}\pm 0.009_{\alpha_s}
\pm 0.003_{\rm exp}\,\mbox{GeV}
$ 
for the bottom 1S mass and
$
\overline m_b(\overline m_b)= 4.235 \pm 0.055_{\rm pert}
\pm  0.003_{\rm exp} \,\mbox{GeV}
$
for the bottom $\msb$ mass, where we have quoted the perturbative error and the
uncertainties from the strong coupling and the experimental data.
\end{abstract}
\maketitle


\newpage

\section{Introduction}
\label{sectionintro}

Determinations of the bottom quark mass $m_b$ have a long history and belong to
the early precision analyses carried out in QCD that required  higher
order computations in the framework of the operator product expansion (OPE)~\cite{
Voloshin:1978hc,Reinders:1981si}. Phenomenologically they are an important input
in the theoretical description of semi-inclusive B-meson decays and
studies of $\Upsilon$-mesons, but they also have some relevance for Higgs
decays into $b$-jets or unification and beyond the standard model studies. Apart
from their phenomenological relevance, 
precision determinations of heavy quark masses are also a reflection of the
ability to do precision computations for the quantities from which they are
extracted, so that continued work in refining the respective theoretical methods
and cross checks with other determination methods are warranted. Since quark
masses are not physical observables due to the confinement, which binds quarks
into hadronic states, they represent renormalization-dependent formal field
theoretic quantities that are defined through the renormalization conditions
imposed to remove the ultraviolet divergences in loop computations. Since the
achievable precision in the bottom quark mass is at the percent level, and thus
substantially smaller than the hadronization scale $\Lambda_{\rm QCD}\sim
200$~MeV, it is mandatory to adopt so-called short-distance mass schemes for 
the bottom quark mass, which are free of ${\cal O}(\Lambda_{\rm QCD})$ infrared
renormalon ambiguities. Short-distance mass schemes involve infrared
subtractions in order to reduce the sensitivity of the perturbative series to
small momenta. Thus short-distance masses either have an intrinsic or an
explicit dependence on the infrared subtraction scale which should be
compatible with the typical short-distance scale governing the heavy quark
mass-dependent observable~\cite{Beneke:1998rk,Pineda:2001zq,Hoang:1999zc}.  

One of the classic methods to determine the bottom quark mass are the $\Upsilon$
sum rules. They involve moments $P_n$  of the total bottom pair production cross
section in $e^+e^-$ annihilation through a virtual photon, defined by 
\begin{align}
P_n \, = \,
\int \frac{d s}{s^{n+1}}\,R_{b\bar b}(s)
\,,
\label{Pndef1}
\end{align}
where $R_{b\bar b}=\sigma(e^+e^-\to b\bar b+X)/\sigma_{\rm pt}$ is the inclusive
bottom pair production cross section and $\sigma_{\rm
  pt}=4\pi \alpha^2/3s$ is the point cross section. 
Conceptually the moments $P_n$ are related to the coefficients of the
zero-momentum expansion of the photon vacuum polarization function coming from
$b\bar b$ quark pairs~\cite{Novikov:1977dq}  
\begin{align}
P_n \,=\, \frac{4\pi^2 Q_b^2}{n! \,q^2} \bigg(\frac{d}{dq^2} \bigg)^n \Pi^\mu_\mu(q)|_{q^2=0}\,,
\end{align}
which can be determined from an OPE where the power
series is dominated by the leading perturbative contribution and the power
corrections represent an expansion in $\Lambda_{\rm QCD}$ over the
short-distance scale $m_b$.  

Since for large values of $n$ the energy range $\Delta E$, where $R_{b\bar b}$ can effectively contribute
to $P_n$ in Eq.~\eqref{Pndef1}, scales as $m_b/n$ \cite{Voloshin:1995sf}, the scale $m_b/n$ emerges as an additional
short-distance scale in the OPE and the values of $n$ one can use for reliable perturbative
analyses are limited up to around $10$. Concerning the methods required for
theoretical computations, two regions in $n$ can be distinguished. For $n\lsim
3$ the theoretical moments are dominated by fluctuations at the scale $m_b$, such
that the usual loop expansion is sufficient to determine the perturbative
contributions. A suitable short-distance mass for the low-$n$ $\Upsilon$ sum
rules is the $\overline{\mbox{MS}}$ mass which has the bottom mass itself as the intrinsic
infrared subtraction scale. On the experimental side, the continuum region for
$R_{b\bar b}$ has a substantial contribution to the moments.  Since there are
sizable energy regions above the $\Upsilon$ bound state regime where there
aren't any experimental measurements, one has to rely on extrapolations for the
experimental moments $P_n^{\rm exp}$. Analyses available in the literature for
the low-$n$ sum rules differ concerning the uncertainty associated to this
extrapolation, see e.g. the analyses in Refs.\cite{Corcella:2002uu} and \cite{Chetyrkin:2010ic}. There are
variants of this method, called finite-energy sum rules where certain energy
ranges are excluded entirely from the analysis~\cite{Bodenstein:2011fv}. For finite-energy sum
rules systematic methods for computing power corrections in the OPE are
currently unknown. 

For $4 \lsim n\lsim 10$ the moments are dominated by the
$\Upsilon$ bound state region and non-relativistic heavy quark pair bound state
dynamics. The large-$n$ sum rules are therefore also called
non-relativistic or $\Upsilon$ sum rules. 
They are the focus of the present work. For the large-$n$ sum rules, in addition
to $m_b$, also $m_b/\sqrt{n}$ and 
$m_b/n$ emerge as additional scales in association to the hard, soft and
ultrasoft momentum scales known from heavy quarkonium physics, and one has to
rely on non-relativistic effective field theory methods based on NRQCD~\cite{Bodwin:1992ye}
to determine the perturbative contributions. Suitable mass definitions belong to
the class of the so-called threshold masses, which have the soft scale
$m_b/\sqrt{n}$ as the intrinsic infrared subtraction scale. Several threshold masses have been defined in the
literature~\cite{Beneke:1998rk,Hoang:1999zc,Hoang:2000yr,Pineda:2001zq}.
Concerning experimental 
data, no extrapolations are required, since the experimentally very well
measured $\Upsilon$ resonances entirely dominate the moments $P_n^{\rm exp}$ and
the continuum contributions of $R_{b\bar b}$ where no measurements exist are
strongly suppressed. For a summary of various bottom mass 
determinations we refer to Ref.~\cite{Beringer:1900zz}.   

In this work we are concerned with the large-$n$ non-relativistic sum
rules. Because of the additional scales, perturbation theory is a simultaneous
expansion in $\alpha_s$ and $1/\sqrt{n}$, where terms proportional to powers of
$(\alpha_s\sqrt{n})$ or $(\alpha_s\ln(n))$ count as order unity and are summed
to all orders. The Coulomb singular terms $\propto (\alpha_s\sqrt{n})^k$ are generated from
ladder-type diagrams with iterated potential interactions and in the so-called
fixed-order expansion only these terms are being summed
systematically within the NRQCD formalism~\cite{Bodwin:1992ye}. Large-$n$ $\Upsilon$ sum
rules analyses were first considered in the fixed-order expansion by
Voloshin in Ref.~\cite{Voloshin:1995sf} at leading-order (LO) and improved up to
next-to-next-to-leading order (NNLO) in
Refs.~\cite{Hoang:1998uv,Melnikov:1998ug,Beneke:1999fe,Hoang:1999ye,Hoang:2000fm} with
differing treatments of the logarithmic 
terms and the residual higher order terms and with different choices of threshold
masses. Unfortunately, the moments in the fixed-order expansion do not exhibit any
good convergence properties and in Refs.~\cite{Hoang:1998uv,Melnikov:1998ug,Beneke:1999fe,Hoang:1999ye} various ad hoc
methods were devised to extract the bottom quark mass at NNLO with a small
uncertainty. We refer to the review~\cite{Battaglia:2003in} for a comparison and
comments. Although
the results of these fixed-order analyses were compatible, the situation was
certainly unsatisfactory. We will comment on results obtained in the fixed-order
expansion in Sec.~\ref{sectiondiscussion}.

In the renormalization group improved (RGI) expansion
also the logarithmic terms are summed systematically with the order counting scheme
\begin{align}
\label{counting}
P_n \,\sim \,\sum_{k,l}(\alpha_s\sqrt{n})^k(\alpha_s\ln(n))^l\,\Big[
1\,\mbox{(LL)};\,
\alpha_s,1/\sqrt{n}\,\mbox{(NLL)};\,
\alpha_s^2,\alpha_s/\sqrt{n},1/n\,\mbox{(NNLL)};\, \ldots
\Big]\,
\end{align}
for the leading logarithmic (LL), next-to leading logarithmic (NLL) and
next-to-next-to leading logarithmic (NNLL) order. The RGI computations can be
carried out in extensions of the original NRQCD formalism. The pNRQCD
formalism~\cite{pNRQCDfirst} is based on the static expansion and factorizes soft and
ultrasoft modes. The vNRQCD formalism~\cite{Luke:1999kz} is based on the correlation of
soft and ultrasoft modes through the energy-momentum dispersion relation for
dynamical heavy quarks. Both formalisms are orthogonal concerning their conceptual
basis and have different Lagrangian formulations, but results which became
available within both formalisms agree.  

Prior to this work a RGI bottom
quark mass determination in the RS-scheme was carried out in
Ref.~\cite{Pineda:2006gx} (see also~\cite{Pineda:2006ri}) at
NLL order. This analysis also contained NNLL order matrix element corrections,
but not the NNLL order corrections to the renormalization group evolution of the
dominant quark pair production current which directly affect the normalization
of the moments $P_n$. The NNLL order corrections to the current anomalous
dimension have a contribution from genuine three-loop diagrams, called
non-mixing terms, and a contribution from the subleading running of the Wilson
coefficients entering the NLL anomalous dimension, called mixing terms. 
The ultrasoft contributions to the subleading running of the relevant coefficients were
determined recently in Refs.~\cite{Hoang:2006ht,Hoang:2011gy} in vNRQCD and in
Ref.~\cite{Pineda:2011aw} in pNRQCD. Although the  
computations differed substantially concerning their conceptual setup and
computational effort, the results from vNRQCD and pNRQCD agree. In
Refs.~\cite{Hoang:2006ht,Hoang:2011gy} also the complete set of the NNLL
ultrasoft mixing contributions to the current anomalous dimension were given.
The complete set of non-mixing terms
has been determined already some time ago in Ref.~\cite{Hoang:2003ns} in vNRQCD, and for
the consistent computation the correlated treatment and the simultaneous
presence of soft and ultrasoft modes was essential. This is due to a term
$\propto\ln(\mu_S^2/m_b\mu_U)$ appearing in the anomalous dimension, which 
enforces the relation $\mu_U\propto\mu_S^2/m_b$ between
the soft renormalization scale $\mu_S$ and the ultrasoft renormalization scale $\mu_U$. 
The corresponding calculations have not yet been achieved in pNRQCD. 
The results for the mixing and non-mixing terms each exhibit very large
corrections from ultrasoft gluons. As was found in Ref.~\cite{Hoang:2011gy}, these large
ultrasoft contributions cancel to a large extent in the sum, confirming a
speculation made in Ref.~\cite{Hoang:2003ns}. Also, the known soft corrections to the current
anomalous dimension, which include the complete set from the non-mixing
terms~\cite{Hoang:2003ns} and the spin-dependent contributions from the mixing
terms~\cite{Penin:2004ay}, are much smaller than the ultrasoft ones and even smaller than  
the typical residual scale dependence of the combined ultrasoft
corrections~\cite{Hoang:2011it}.

In this work we determine the 1S~bottom quark mass~\cite{Hoang:1999zc} from the large-$n$
moments $P_n$ in the RGI expansion including all known NNLL order corrections,
in particular the new contributions to the Wilson coefficient of the
dominant quark pair production current. Our analysis is based on fits of single
moments, and we examine in detail the convergence properties of the moments and
their perturbative uncertainties. For the still unknown NNLL soft mixing
contributions to the anomalous dimension of the dominant quark pair
production current we give arguments that their size is comparable to
the known soft contributions. The resulting uncertainty is tiny, and in
particular smaller than 
the remaining scale variation of the known ultrasoft corrections. We therefore
consider our analysis as NNLL order. Compared to the situation in the fixed-order
expansion we find a reasonably good behavior of the perturbation series which
does not require any ad hoc methods to determine the bottom quark mass.

The outline of this paper is as follows:
In Sec.~\ref{sectionnewtheory} we discuss the theoretical input for the
non-relativistic factorization formula for the NNLL
order moments we use for our analysis. For the most part we review older
results, but we also discuss and numerically analyze the new NNLL order
corrections to the anomalous dimension of the leading quark pair production
current. An important aspect of the numerical analysis is that these NNLL order
corrections are sizable. In Sec.~\ref{sectionpertexp} we therefore analyze the
perturbative expansion and find that only fully expanding out all terms in the
NNLL order factorization formula leads to the desired non-relativistic
renormalization scaling behavior. Our numerical analysis concerning fits for the
bottom mass determination and the details on how we implement renormalization
scale variations and estimate the perturbative uncertainty are discussed in
Sec.~\ref{sectionanalysis}. 
In Sec.~\ref{sectiondiscussion} we discuss the results of the fits when we
modify our moments to account only for the terms required in the fixed-order
expansion and demonstrate the improvement achieved by the resummation of logarithms. 
We also present arguments that disfavor carrying out simultaneous
fits with several moments using a $\chi^2$ function due to the strong
correlation between different large-$n$ moments. Section~\ref{sectionconclusion}
contains the conclusions. 
We have also added three appendices giving details on the analytic results for
NNLL order moments and presenting the experimental moments.

\section{Theory Moments with RG Improvement}
\label{sectionnewtheory}

For the determination of the theoretical expressions for the large-$n$ moments
with renormalization group improvement we follow closely the approach of
Refs.~\cite{Voloshin:1995sf,Hoang:1998uv,Hoang:1999ye} to implement the
simultaneous expansion in $\alpha_s$ 
and $1/\sqrt{n}$ according to the scheme in Eq.~(\ref{counting}), and, here,
mainly concentrate on explaining the outline of the calculations and the new
aspects that arise for the RG-improved calculation. 
The NNLL order large-$n$ moments can be determined from the integral 
\begin{align}
\label{Pnexpression1}
P_n^{th,\rm NNLL} &=
\frac{1}{4^{n} (M_b^{pole})^{2n}} \! \int\limits_{E_{\rm bind}}^\infty \!\!
\frac{d E}{M_b^{pole}} \exp\Big\{\!\!
-\frac{E}{M_b^{pole}} \,n \Big\}
\Big( 1 - \frac{E}{2\,M_b^{pole}} + \frac{E^2}{4\,(M_b^{pole})^2}\,n \Big)R_{\rm NNLL}^{\rm thr}(E)
\,,
\end{align}
where $E\equiv\sqrt{s}-2M_b^{pole}$ is the energy w.r.t. the pole mass
threshold and $E_{\rm bind}$ is the binding
energy of the lowest lying 1S resonance. 
Compared to the original definition of Eq.~(\ref{Pndef1})
the weight function $1/s^{n+1}$ has been rewritten as an exponential expansion that
explicitly separates the LL contributions from the NNLL order kinematic
corrections. For the normalized $b\bar b$ total cross section $R$-ratio in the
threshold region at NNLL order, $R^{\rm thr}_{\rm NNLL}$, one has
to insert the absorptive part of the leading order and $O(v^2)$-suppressed
non-relativistic vector current correlators arising for the heavy quark pair
production process through a virtual photon in $e^+e^-$ annihilation. The
required results are well known from the literature on top pair production in the
threshold region at a future linear collider and can be applied to the bottom
pair cross section with trivial adaptations mostly related to the different
number of active flavors. The explicit results can be taken 
from Ref.~\cite{Hoang:2000ib,Hoang:2001mm,Hoang:2002yy} supplemented by the
modifications given in Ref.~\cite{Hoang:2003ns} 
arising from a more convenient convention for the $1/(m k)$-type
potentials. The contributions to the moments related to the
Coulomb interaction are the same as for the fixed-order moments up to the additional
logarithms summed 
in the leading order Coulomb potential Wilson coefficient and have been given
explicitly in Ref.~\cite{Hoang:1998uv}. Deforming the energy integration into
the complex plane the integral in Eq.~\eqref{Pnexpression1} can be quite conveniently carried out by inverse Laplace
transform methods. The corresponding formulae can be 
easily read off from tables and have been collected for convenience in
Ref.~\cite{Hoang:1998uv}. 

The NNLL expression for the n-th moment can then be cast into the form
\begin{align}
P_n^{th,\rm NNLL} &=
\frac{3\,N_c\,Q_b^2\,\sqrt{\pi}}{4^{n+1} (M_b^{\rm pole})^{2n} \, n^{3/2}}\,
\bigg\{\,
c_1(h,\nu)^2\,\varrho_{n,1}(h,\nu) +
2\, c_1(h,\nu) c_2(h,\nu) \,\varrho_{n,2}(h,\nu)
\,\bigg\}\,,
\label{Pnth}
\end{align}
where $\varrho_{n,1}$ arises from the integration over the non-relativistic
current correlator involving the dominant effective S-wave current and
$\varrho_{n,2}$ from the current correlator involving one insertion of the
${\cal O}(v^2)\sim{\cal O}(1/n)$-suppressed S-wave current. Both correlators are referred
to as ${\cal{A}}_1$ and ${\cal{A}}_2$, respectively, in
Refs.~\cite{Hoang:2001mm,Hoang:2002yy}. 
The contributions to
$\varrho_{n,1}$ coming from the Coulomb potential are UV-finite and therefore
identical to the results given in Ref.~\cite{Hoang:1998uv} up to the previously
mentioned trivial modification due to the Wilson coefficient of the Coulomb potential which
differs from the strong coupling at NNLL
order~\cite{Pineda:2000gza,Hoang:2001rr,Hoang:2002yy}. The other contributions to 
$\varrho_{n,1}$ as well as the result for $\varrho_{n,2}$, which contribute
exclusively at NNLL order, differ due to different conventions for the
potentials used in Refs.~\cite{Hoang:2002yy,Hoang:2003ns} and also because they involve UV-divergences renormalized in the
$\msb$ scheme (compared to a cutoff scheme used in
Ref.~\cite{Hoang:1998uv}). All expressions for the non-Coulomb terms in
$\varrho_{n,1}$ and for $\varrho_{n,2}$ have not been published before and are
given explicitly in App.~\ref{Apprhos}. 
Throughout this paper all couplings and Wilson coefficients are understood to be renormalized in the $\msb$ scheme.

The terms $c_1$ and $c_2$ are the Wilson coefficients of the dominant and the
${\cal O}(v^2)$ subleading currents, respectively. The NNLL order anomalous dimension of $c_1$
contains the previously mentioned mixing and non-mixing corrections according to
Refs.~\cite{Hoang:2006ht,Hoang:2011gy,Pineda:2011aw} and
Ref.~\cite{Hoang:2003ns}, respectively. These two types of NNLL order
corrections to $c_1$ have not yet been analyzed together in any phenomenological
analysis of bottom pair production close to the threshold. A brief numerical
analysis of $c_1$ focusing on the NNLL order scale variation and the uncertainty
due to the yet unknown soft mixing contributions in the anomalous dimension of
$c_1$ is carried out at the end of this section. Details on the analytic 
results for $c_1$ and $c_2$ are given at the end of this section and in
App.~\ref{Appc1}.

We note that in our analysis we treat the charm quark as
massless. Up to now the effects of the non-zero charm mass have only been
treated in the fixed-order formalism exhibiting a shift in the bottom mass
between $-20$ and $-30$~MeV~\cite{Hoang:1999us,Hoang:2000fm}. The corrections
from the non-zero charm mass $m_c$ are conceptually interesting since $m_c$ is
numerically close to the soft scale $\sim \alpha_s m_b$ and generically above
the ultrasoft scale $\sim\alpha_s^2 m_b$. The effects are therefore not
suppressed by additional factors of $m_c^2/m_b^2$ and are even somewhat enhanced  
since the charm mass interferes as a physical infrared scale in the renormalon
cancellation. At this time a 
systematic renormalization group improved treatment of massive virtual quark 
thresholds in heavy quarkonium production is still lacking. However, taking the
known effects in the fixed-order approach as a guideline, we conclude that the
associated uncertainty is smaller 
than the perturbative uncertainties of our present analysis in
Sec.~\ref{sectionanalysis}. 

The correlators and Wilson coefficients all
depend on the dimensionless vNRQCD velocity renormalization evolution parameter $\nu$
that is used to implement the correlated evolution of soft ($\mu_S$) and
ultrasoft scales ($\mu_U$) according to the non-relativistic scaling
$\mu_U\propto\mu_S^2/m_b$. 
For the large-$n$ moments the typical choice for $\nu$ is 
of order of the velocity of the bottom quarks which generically scales as
$1/\sqrt{n}$.   
The residual dependence of predictions on changes of $\nu$
are used to estimate the perturbative uncertainty of the low-energy
contributions contained in the theoretical prediction.
On the other hand, changes coming from the variation of the matching scale
$\mu_h$ can be used as an estimate of 
the perturbative uncertainty of the high-energy contributions contained in the
moments. Both types of uncertainties can be considered as independent, because
hard and low-energy contributions are separated in the non-relativistic
effective theory. In order to avoid that the 
ultrasoft scale can exceed the soft scale we use the assignment
$\mu_S=\nu \mu_h$, 
$\mu_U=\nu^2\mu_h$
and impose the restriction $\nu\le
1$. Moreover we parametrize the matching scale as $\mu_h= h\, m_b$, where the
typical choice for $h$ is of order one. The renormalization scaling parameter
$h$ has also been indicated in Eq.~(\ref{Pnth}). 

In our parametrization with the
dimensionless renormalization scaling parameters $\nu$ and $h$ all explicit
dependence on the bottom quark mass $m_b$ cancels from the Wilson coefficients
as well as from the matrix elements corrections contained in $\varrho_{n,1}$ and
$\varrho_{n,2}$. The only explicit dependence on the bottom quark mass arises
from the overall dimensional factor shown in Eq.~(\ref{Pnth}). In addition,
there remains an implicit dependence on the bottom mass through the dependence
of the Wilson coefficients or the matrix element corrections on the
strong coupling $\alpha_s$, which has to be evaluated either at the matching, the
soft or the ultrasoft renormalization scales. 
Because the matching scaling
parameter $h$ always occurs together with an additional $m_b$ factor and because the
solutions of the anomalous dimensions can be expressed in terms of the strong
coupling $\alpha_s$ evaluated at the hard, soft or ultrasoft scales, it is
straightforward to implement the $h$-dependence of the moments by simply
inserting a factor of $h$ to the argument of each $\alpha_s$ that arises.     
This method covers all $h$-dependence except for two terms, one arising from the term
$\ln(\mu_S^2/m_b \mu_U)=\ln h$ in the NNLL
non-mixing anomalous dimension of $c_1$~\cite{Hoang:2003ns} and one in the NNLL order matching
condition $c_1(\mu_h,1)$. Both terms are given in App.~\ref{Appc1} as well.

The moments shown in Eq.~(\ref{Pnth}) have been written in terms of the bottom quark pole
mass $M_b^{\rm pole}$, which is known to be sensitive to renormalon long-distance
effects that seriously deteriorate the convergence of the perturbative
expansion. It is therefore mandatory to employ a better defined short-distance
bottom quark mass scheme. Short-distance mass schemes have an intrinsic
dependence on an infrared cutoff scale~\cite{Hoang:2008yj}, and for problems involving
non-relativistic heavy quarkonium dynamics one has to employ so-called threshold
masses, where this scale is of order of the inverse Bohr radius, ${\cal
  O}(m_b\alpha_s)$~\cite{Hoang:2000yr}. In this work we will use the 1S mass
defined as half the 
perturbative contribution to the ${}^3S_1$ bottonium ground state,
$\Upsilon(1S)$~\cite{Hoang:1998ng,Hoang:1998hm}. Other possible threshold mass definitions
have been discussed e.g.\ in Ref.~\cite{Hoang:2000yr}, and in Ref.~\cite{Pineda:2006gx}
the so-called renormalon-subtracted mass scheme was employed.
The result can finally be converted to the 
$\msb$ mass, which is frequently used for predicting physical quantities
involving scales larger than the bottom mass. Although this conversion avoids
the renormalon problem of the pole mass, we note, however, that the
conversion introduces a rather large dependence on the uncertainty in the value
of the strong coupling $\alpha_s$ due to a term $\propto\alpha_s m_b$ appearing
in the conversion series. This motivates quoting the result
for the 1S mass as the main result of our analysis.\footnote{We refer the reader
to Ref.~\cite{Bethke:2011tr} for a review on the in our view incoherent
situation concerning the value and uncertainty in $\alpha_s(M_z)$.} We note,
however, that in our case this $\alpha_s$-dependence just compensates for the
$\alpha_s$-dependence of the 1S mass we determine from our sum rule analysis. 

To convert Eq.~(\ref{Pnth}) to the 1S bottom mass scheme its relation to the pole
mass is required, which at NNLL order reads
\begin{align}
M_b^{\rm pole} &= M_b^{1S} \{1+\Delta^{\rm LL} + \Delta^{\rm NLL} + [(\Delta^{\rm LL})^2 + \Delta_c^{\rm NNLL} + \Delta_m^{\rm NNLL}]\}\,.
\label{mpoletom1s}
\end{align}
The various $\Delta$ terms are given in Ref.~\cite{Hoang:2001mm}, where it is
necessary to use Eq.~(\ref{phi}) for the coupling parameter $a$ and the result
for ${\cal V}_{k,{\rm eff}}^{(s)}$ from Ref.~\cite{Hoang:2002yy} for the
coefficient ${\cal V}_{k}^{(s)}$. 
The logarithms in the $\Delta$ terms are to be understood as
 $L \equiv \ln[ \mu_S/(a m_b) ]$, which reduces to $\ln[ \nu/a ]$ 
for $h=1$ as given in Ref.~~\cite{Hoang:2001mm}.
On the RHS of Eq.~(\ref{mpoletom1s}) the terms
$\Delta^{\rm LL}$ and $\Delta^{\rm NLL}$ are the LL and NLL order contributions in the
non-relativistic expansion, and the three terms in the brackets are the NNLL
order corrections.
As discussed in detail in Ref.~\cite{Hoang:1999ye}, it is crucial to reexpand the
series in a way that is consistent with the exponential non-relativistic
expansion given in Eq.~(\ref{Pnexpression1}) where the LL order contribution to
the binding energy is treated as a leading order term. This is achieved by using
the expression 
\begin{align}
\label{mpoletom1s2}
\frac{1}{(M_b^{\rm pole})^{2n}} &= \frac{1}{(M^{1S}_b)^{2n}} \exp(-2 n \Delta^{\rm LL} ) \nn\\
&\quad \times \Big\{1- 2n\Delta^{\rm NLL} + n \big[ (\Delta^{\rm LL})^2 -2 \Delta^{\rm NNLL} + 2n (\Delta^{\rm NLL})^2 \big] + \ldots \Big\}
\,,
\end{align}
where the first, second and third term in the second line correspond to LL, NLL and NNLL order, respectively. 
In order to consistently remove the infrared renormalon problem associated with
the pole mass when switching to the 1S scheme, it is important to consistently
expand out the series in the curly brackets with the  
corresponding LL, NLL and NNLL order terms contained in $\rho_{n,1}$, while the
exponential term containing $\Delta^{\rm LL}$ remains unexpanded. For the
$\rho_{n,2}$ only the LL 
exponential term in Eq.~(\ref{mpoletom1s2}) has to be employed. Note that it is also
crucial to use the same values for the renormalization scale parameters $\nu$
and $h$ in the Wilson coefficients and couplings of Eqs.~(\ref{Pnth}) and
(\ref{mpoletom1s2}). 
Unless stated explicitly we always refer to the 1S mass definition as the bottom
quark mass: $m_b \equiv M_b^{1S}$.

The contributions to the renormalization group evolution of $c_1$, the Wilson 
coefficient of the 
leading bottom pair production current, can be parametrized as 
\begin{align}
\label{c1solution}
 \frac{c_1(h,\nu)}{c_1(1,1)}& =
\exp\Big[\,\xi^{\rm NLL}(h,\nu) + 
\Big(\,\xi^{\rm NNLL}_{\rm m}(h,\nu) + \xi^{\rm NNLL}_{\rm nm}(h,\nu)\,\Big) 
+ \ldots\,\Big]
\\\nonumber
& = 1+ \xi^{\rm NLL} + 
\Big[ \Big(\frac{1}{2}\xi^{\rm NLL}\Big)^2 + \xi_{\rm m}^{\rm NNLL}+\xi^{\rm NNLL}_{\rm
  nm}\Big]
+ \ldots\,
\,,
\end{align}
where $\xi^{\rm NLL}$ refers to the NLL order contribution and the $\xi^{\rm
  NNLL}_{\rm m}$ and $\xi^{\rm NNLL}_{\rm nm}$ to the NNLL order mixing and
non-mixing corrections, respectively. The matching condition $c_1(h,1)$ and the
expressions for the $\xi$'s can be
found in App.~\ref{Appc1} and the references cited there. In the second line of
Eq.~(\ref{c1solution}) we have displayed the consistently expanded form where
the terms in the squared brackets represent the NNLL corrections
and the explicit dependence of the $\xi$'s on the scale parameters 
has been suppressed. 
We note that since the
LL order anomalous dimension for the current is zero, the expanded form does not
contain any exponentiated LL order contribution.

Based on the recently completed calculation of the ultrasoft NLL running of the
Wilson coefficients associated to the $\ord(v^2)$- and $\ord(\alpha_s v)$-suppressed 
potentials~\cite{Pineda:2011aw,Hoang:2011gy,Hoang:2006ht}, 
the ultrasoft mixing contributions to $\xi^{\rm NNLL}_{\rm m}$, referred to as
$\xi^{\rm NNLL}_{\rm m, usoft}$ has been determined in Ref.~\cite{Hoang:2011gy}.
Concerning the corresponding soft mixing contributions currently only those coming from
the NLL order anomalous dimension of the spin-dependent $\ord(v^2)$-suppressed
potential are fully known and were found to be tiny~\cite{Penin:2004ay}.
On the other hand, because the NLL matching conditions of all
the suppressed potentials are known, it is possible to compute the term 
$\propto\alpha_s^3\ln\nu$ of the soft mixing contributions, which we call 
$\xi^{\rm  NNLL}_{\rm m, soft 1}$. The result was already presented in
Ref.~\cite{Hoang:2003ns} and it also given in Eq.~(\ref{softmixlog}).

The NNLL order non-mixing contributions in $\xi^{\rm NNLL}_{\rm nm}$, 
i.e. the ultrasoft and soft contributions, 
are both completely known already since some time
from Ref.~\cite{Hoang:2003ns}. 
In the same publication it was observed that the
ultrasoft non-mixing contributions were more than an order of magnitude larger
than the soft ones, and that the smallness of the latter was not arising from any
accidental cancellation between different color factors but was a genuine
property of all soft non-mixing contributions. 
In the following we analyze all known NNLL contributions together numerically
and give arguments suggesting that the ultrasoft contributions also dominate the
mixing corrections and that the unknown soft corrections are negligible.

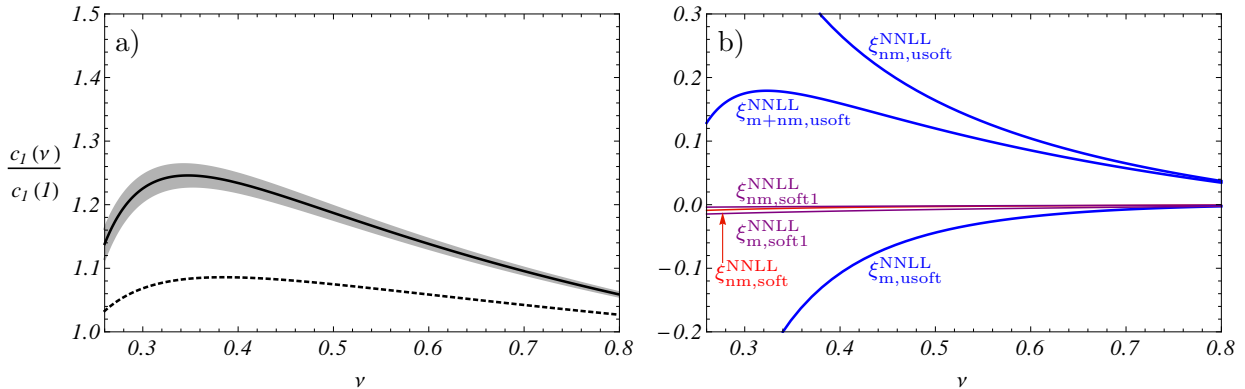
\begin{figure}[th]
\includegraphics[width=1.02\textwidth]{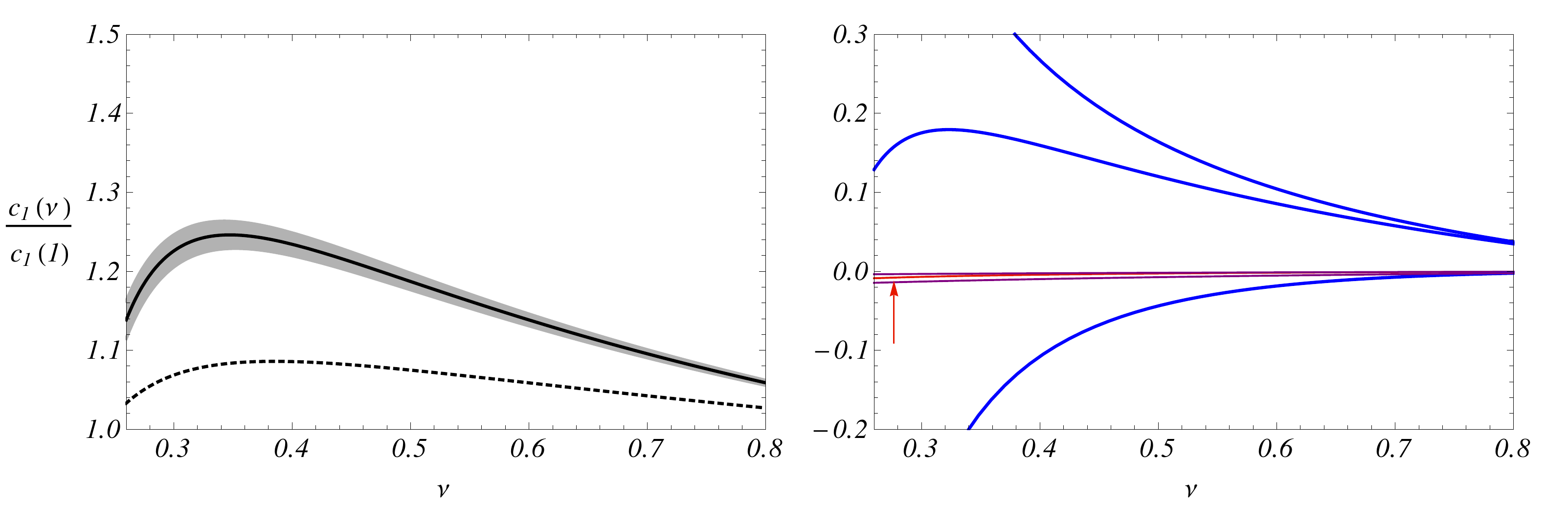}
\put(-435,135){a)}
\put(-207,135){b)}
\put(-150,135){\scriptsize \blau $\xi_{\rm nm,usoft}^{\rm NNLL}$}
\put(-150,50){\scriptsize \blau $\xi_{\rm m,usoft}^{\rm NNLL}$}
\put(-200,110){\scriptsize \blau $\xi_{\rm m+nm,usoft}^{\rm NNLL}$}
\put(-200,80){\scriptsize \lila $\xi_{\rm nm,soft1}^{\rm NNLL}$}
\put(-200,64){\scriptsize \lila $\xi_{\rm m,soft1}^{\rm NNLL}$}
\put(-208,49){\scriptsize \rot $\xi_{\rm nm,soft}^{\rm NNLL}$}
\caption{Panel a): 
RG evolution of the  ${}^3S_1$ current coefficient
$c_1(\nu)$ normalized to $c_1(1)$ for $m_b=4.7$~GeV and $h=1$. 
The dashed line represents the full NLL result in expanded form: $1+\xi^{\rm NLL}$.
The solid line accounts for all known contributions including the new ultrasoft
NNLL mixing correction in Eq.~\eqref{XiNNLLmixusoft} as well as the leading soft
mixing logarithm at NNLL $\xi^{\rm NNLL}_{\rm m, soft1}$ in
Eq.~\eqref{softmixlog}: $1+ \xi^{\rm NLL} +(\xi^{\rm NLL})^2/2 + \xi_{\rm
  nm}^{\rm NNLL}+\xi^{\rm NNLL}_{\rm m, usoft}+\xi^{\rm NNLL}_{\rm m, soft1}$.  
The gray area around the black line is generated by varying the known soft NNLL
contributions to that curve by a factor between 0 and 2. 
Panel b):
Separate curves for the different (soft/ultrasoft, mixing/nonmixing) NNLL
corrections ($\xi^{\rm NNLL}$) to the running of $c_1$ as indicated in the
plot. 
The NNLL contribution coming from the exponentiation of the NLL term, $(\xi^{\rm
  NLL})^2/2$, is tiny. Since it virtually coincides with the horizontal zero-line
in this plot, it has not been drawn. The curve for $\xi_{\rm nm,soft1}^{\rm
  NNLL}$ represents the linear logarithmic term $\propto\alpha_s^3\ln\nu$
contained in the complete soft non-mixing correction $\xi_{\rm nm,soft}^{\rm
  NNLL}$.  
For both plots we have used $\alpha_s^{(n_f=4)}(4.7~\mbox{GeV})=0.217$.
\label{c1Plot}} 
\end{figure}
In the left panel of Fig.~\ref{c1Plot} the renormalization scale evolution
of $c_1(h,\nu)/c_1(1,1)$ is displayed for $0.26 \le\nu\le
0.8$ at NLL order (dashed line) and including all known NNLL order
contributions, where the soft mixing corrections are approximated by 
$\xi^{\rm NNLL}_{\rm m, soft 1}$ (solid line).
The gray band drawn around the NNLL result arises from the variation of the
result coming from multiplying the soft terms at NNLL order by a factor
between zero and two. Comparing the width of the gray band with the overall
shift between the NLL and NNLL order results and also with the overall
remaining $\nu$ dependence of $c_1(h,\nu)/c_1(1,1)$ at NNLL order we make two
important observations. 

First, the ultrasoft contributions dominate the
NNLL order result and are about an order of magnitude larger than the known soft
contributions. From the right panel of Fig.~\ref{c1Plot} we see that this is
true for the mixing and non-mixing contributions separately as well as for the
sum. From the close vicinity of the curves for the full expression for  
$\xi_{\rm nm,soft}^{\rm NNLL}$ and its linear logarithmic contribution
$\xi_{\rm nm,soft1}^{\rm NNLL} \propto \alpha_h^3 \ln\nu$, given in
Eq.~\eqref{softnonmixlog}, we can also conclude 
that the curve for $\xi^{\rm NNLL}_{\rm m, soft1}$ should represent the correct
order of magnitude for the full set of soft mixing corrections.
It is therefore reasonable to take the gray band as an uncertainty associated to
the currently unknown soft mixing corrections. Comparing the gray band to the
size and scale dependence of the bigger ultrasoft corrections we can conclude that
this uncertainty is smaller than the remaining NNLL renormalization scale variation
we observe in the relevant range $0.3\lesssim\nu\lesssim 0.6$. This suggests that the unknown soft
corrections can be 
neglected for a NNLL order prediction. 

As the second observation, we see that
the NNLL order corrections to the anomalous dimension of $c_1$ are about a
factor two larger than the NLL order ones. This somewhat unsettling fact might
cast slight doubts concerning the quality of the perturbative expansion due to
large ultrasoft corrections, but
might be as well associated to an anomalously small NLL order correction, which
amounts to less than $10\%$. Nevertheless, given the situation it is certainly
appropriate to invest
a closer look on the behavior of the perturbation series for the moments.
  
Finally we comment on the presence of nonperturbative power corrections in
the OPE of the current correlators ${\cal{A}}_i$. The effect of the leading
power correction on the theoretical large-$n$ moments in Eq.~\eqref{Pnth} is
associated with the ${\cal O}(\lqcd^4)$ gluon condensate arising in the OPE for
${\cal{A}}_1$. 
The corresponding nonperturbative power correction to 
the $n$-th moment can be written approximately
as~\cite{Voloshin:1995sf} 
\begin{align}
\delta P_n^{th,np} &=  P_n^{th, \rm LL} \, \langle \alpha_s\, {\bf G^2} \rangle
\frac{n^3 \pi}{72\, m_b^4} \exp(-0.4\, C_F \alpha_s \sqrt{n})\,,
\label{np}
\end{align}
where we take $\alpha_s$ in the exponent at the soft scale. 
To our knowledge the subleading corrections to the Wilson coefficient 
of the gluon condensate are currently unknown. Numerically
the relative correction from Eq.~\eqref{np} is less than a percent for $n<20$,
i.e.\ even smaller then the experimental uncertainty and therefore negligible for
our purpose, cf. Sec.~\ref{sectionanalysis}. In order to make sure that also
higher order power corrections~\cite{Pineda:1996uk} are sufficiently suppressed,
we, however, only consider $n \lesssim 10$ for our analysis.

\section{The perturbative expansion}
\label{sectionpertexp}

The perturbative series for the large-$n$ moments $P^{th}_n$ 
in Eq.~\eqref{Pnth} follows
the non-relativistic expansion scheme of Eq.~(\ref{counting}) and, as explained
in the previous section, requires that
the perturbative series for $\rho_{n,1}$ coming from the leading order current
correlator is consistently expanded with the additional corrections of
Eq.~(\ref{mpoletom1s2}) that arise from employing the 1S threshold mass scheme.
However, concerning the factor $c_1^2$ in Eq.~\eqref{Pnth} one has the option either to
keep it as an overall factor for the series associated to $\varrho_{n,1}$
(``factorized form'') or to
expand it out as well together with the series of $\varrho_{n,1}$ (``expanded
form''). The difference between the two ways to expand $P^{th}_n$ represents terms
from beyond NNLL order and should, in principle, be within the theoretical
uncertainty of our NNLL result. However, given the observations concerning the
perturbative behavior of $c_1$ discussed above we believe that it is mandatory
to have a closer look on the expanded and factorized expansions of the moments.

An important property of the large-$n$ moments is that the non-relativistic
dynamics encoded in them is governed by bottom quark velocities $v\sim {\cal
  O}(1/\sqrt{n})$. Since the vNRQCD renormalization scaling parameter $\nu$ is
typically of order $v$, we would expect that a good convergence of the series
and the best physical description is
achieved for $\nu\sim (1/\sqrt{n}+\text{const.})$, where the additional constant
term arises from a saturation effect that protects the renormalization scale
from vanishing for large $n$.\footnote{As argued in Ref.~\cite{Voloshin:1995sf}
  the proper choice for the soft ($\mu_S$) and ultrasoft renormalization scale
  ($\mu_U$) in the limit $n\to \infty$ approaches the inverse Bohr radius ($\sim
  \alpha_s m_b$) and the binding energy ($\sim \alpha_s^2 m_b$) of the lowest lying
  resonance, respectively.}   
Next we examine for the expanded and the factorized
form of the large-$n$ moments, how well they satisfy this behavior. 

\begin{figure}[th]
\begin{center}
\includegraphics[width=0.48\textwidth]{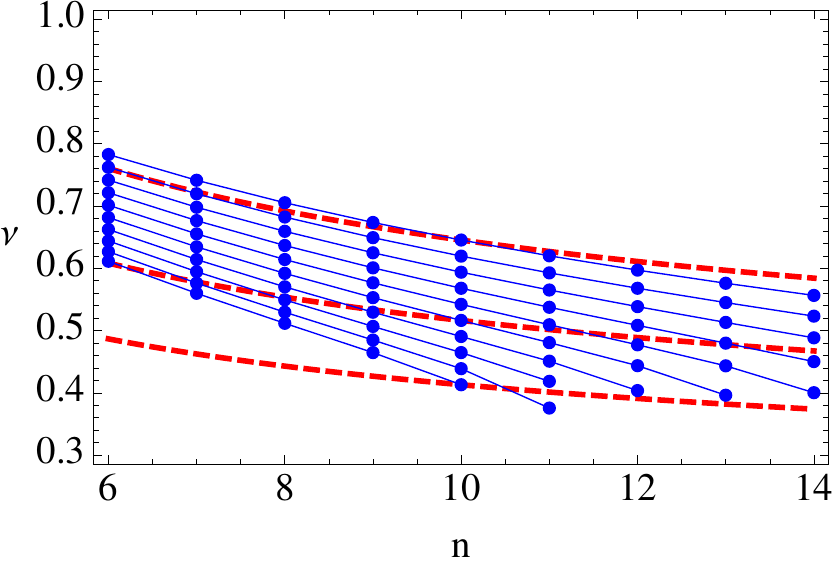}
\hfill
\includegraphics[width=0.48\textwidth]{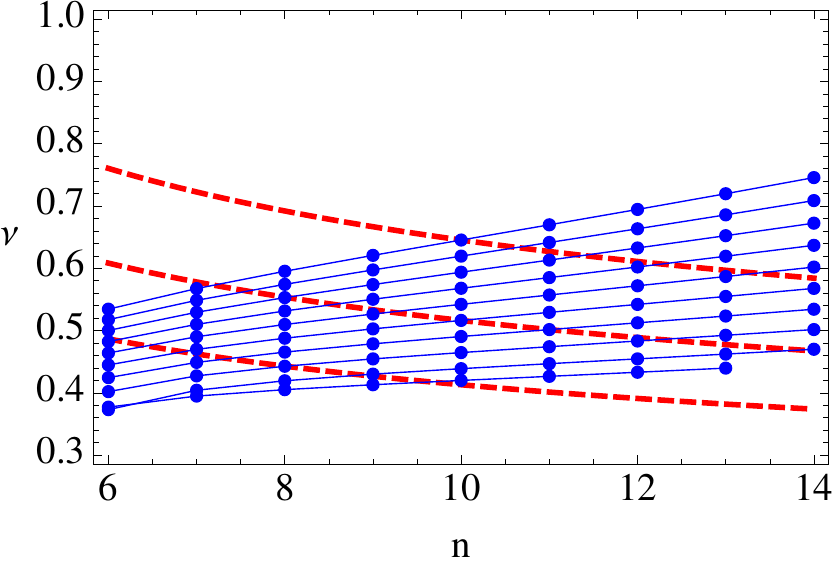}
\put(-437,138){a)}
\put(-194,138){b)}
\end{center}
\caption{Values for the renormalization parameter $\nu$ required by
  $P^{th}_n(m_b;\nu) = P^{exp}_n$ for fixed mass $m_b$ and different $n$. The
  blue lines connect the $\nu$ values (blue dots) for the same reference
  mass. Lines for 10 different reference masses determined from
  $P^{th}_{10}(m_b; \nu\!=\!f(1/\sqrt{10} + 0.2)) = P^{exp}_{10}$ with $f$
  between 0.8 and 1.25 are shown. For the plot in panel $a)$ we used the fully
  expanded NNLL expression and for the plot in panel $b)$ the factorized NNLL
  expression for $P^{th}_n$, as described in the text. 
The reference masses in a) range from $4.77$~GeV for $f=0.8$ to $4.74$~GeV for $f=1.25$.
The reference masses in b) range from $4.69$~GeV for $f=0.8$ to $4.64$~GeV for $f=1.25$.
The dashed red curves are defined by $\nu = f(1/\sqrt{n} +0.2)$ with
$f=0.8,1,1.25$ respectively. 
\label{nuovern}}
\end{figure}

In Fig.~\ref{nuovern} we plot the values of the renormalization parameter $\nu$
that are required to satisfy $P^{th}_n(m_b;\nu) = P^{exp}_n$ as a function of
$n$ using $h=1$ and a fixed value of $m_b$, where the
$P^{exp}_n$ are the experimental moments. The value of the bottom quark
mass has been fixed by fitting the theoretical $10$-th moment, $P^{th}_{10}$, to
the experimental one using $\nu=f(1/\sqrt{10}+0.2)$. Here the constant $f$ is a
fudge parameter which should be of order one and which we have varied between
$0.8$ and $1.25$ in steps of $0.05$. The results for $\nu$ we obtained for
$n\neq 10$ are shown as blue dots, and values for equal $f$ are connected by the
blue lines. The left panel shows the outcome for the expanded moments and the
right panel for the factorized moments. We see that the $\nu$ values for the
fully expanded moments decrease with $n$ and are indeed consistent with the
expected behavior $\nu\sim (1/\sqrt{n}+\mbox{const.})$. On the other hand, the
$\nu$ values from the factorized moments increase with $n$ and are inconsistent
with the expected behavior. As a comparison we have displayed the
functions  $\nu=f(1/\sqrt{n}+0.2)$ with $f=0.8,1.0,1.25$ as the dashed red lines.

The quite different outcome for the expanded and factorized moments is caused by
the perturbative behaviour of $c_1$ discussed in the previous section. 
This shows that the (expansion scheme dependent) fraction of terms from beyond
the NNLL level included in the two ways of the NNLL expansion for the moments,
are large enough to qualitatively affect renormalization scaling features of
the perturbative result. 
Because only the fully expanded moments show the expected behavior for their
renormalization scaling parameter $\nu$, we adopt them for our numerical
analysis to determine the bottom quark mass. However, we require as an important
cross check that the bottom mass values determined from the factorized moments
have to be consistent with those from the expanded moments within the
perturbative uncertainty.

\section{Numerical Analysis}
\label{sectionanalysis}

The experimental moments we use for our analysis are based on the measurements
of the masses and electromagnetic decay widths of the first four $\Upsilon$
resonances ($\Upsilon(1S)$-$\Upsilon(4S)$)~\cite{Beringer:1900zz} in the narrow width
approximation, BABAR data for the threshold 
region~\cite{Aubert:2008ab} in the energy range between $\sqrt{s}=10.62$ and
$\sqrt{s}=11.21$ and the perturbative QCD result for the continuum at energies
$\sqrt{s} > 11.21$~\cite{Chetyrkin:1997pn}, where no experimental data is
available.\footnote{We thank V.~Mateu and B.~Dehnadi for providing us the data
  compilation and a numerical code for the experimental moments.}  
As we 
are aiming at an analysis for large-$n$ 
moments, the $\Upsilon$ resonances and the BABAR region constitute the major
part of the experimental moments (($87$,$93$,$96$,$98$)\% for
  for the $\Upsilon$ resonances, and ($5.7$,$3.7$,$2.3$,$1.4$)\%
  for the BABAR region  for $n=(6,8,10,12)$) and only little effort has to be
invested for the continuum region above 
$11.21$~GeV where no experimental data exists. We follow the approach of
Ref.~\cite{Dehnadi:2011gc} for the BABAR region and also adopt an additional
$10\%$ model uncertainty for the contiuum 
region. We stress, however, that the latter uncertainty only constitutes a very
small and numerically irrelevant part
of the whole error budget of the large-$n$ moments which are dominated by the
uncertainties from the $\Upsilon$ resonances and the BABAR region.  
We note that applying our approach for low-$n$ moments, we obtain results that
are perfectly compatible with the moments given in Ref.~\cite{Kuhn:2007vp}
albeit with somewhat larger errors if applied for low $n$ values, as in their
approach only perturbative QCD 
uncertainties were accounted for in the continuum region. The mean values
together with the respective statistical and systematic errors of the
$P^{exp}_n$ for $4 \le n \le 14$ are given in App.~\ref{AppPexp}. 

In order to determine the 1S bottom mass from single moment fits we simply
solve the equation $P^{th}_n(m_b) = P^{exp}_n$ for $m_b$. In order to estimate
the perturbative error we carry out these fits many times varying matching and
renormalization scales in reasonable ranges. To parametrize the scale variations
for the different moments $P^{th}_n$ such that the $n$-dependence of the
renormalization scales and their correlation is implemented coherently 
we use the assignments 
\begin{align}
\label{scales}
\mu_h=h\, m_b\,, 
\mu_S=h\, m_b f \nu^*
\quad\mbox{and}\quad 
\mu_U =h\, m_b (f
\nu^*)^2
\end{align}
 for the matching, soft and ultrasoft scales, respectively, where
$\nu^* = (1/\sqrt{n} +0.2)$ is our default choice for the 
renormalization scaling parameter $\nu$. So the matching, soft and ultrasoft
scales adopt their default values for $h=f=1$, and we parametrize variations of
these scales by scanning over a certain region in the two-dimensional
$h$-$f$-plane.   

\begin{figure}[t]
\begin{center}
\includegraphics[width=0.5\textwidth]{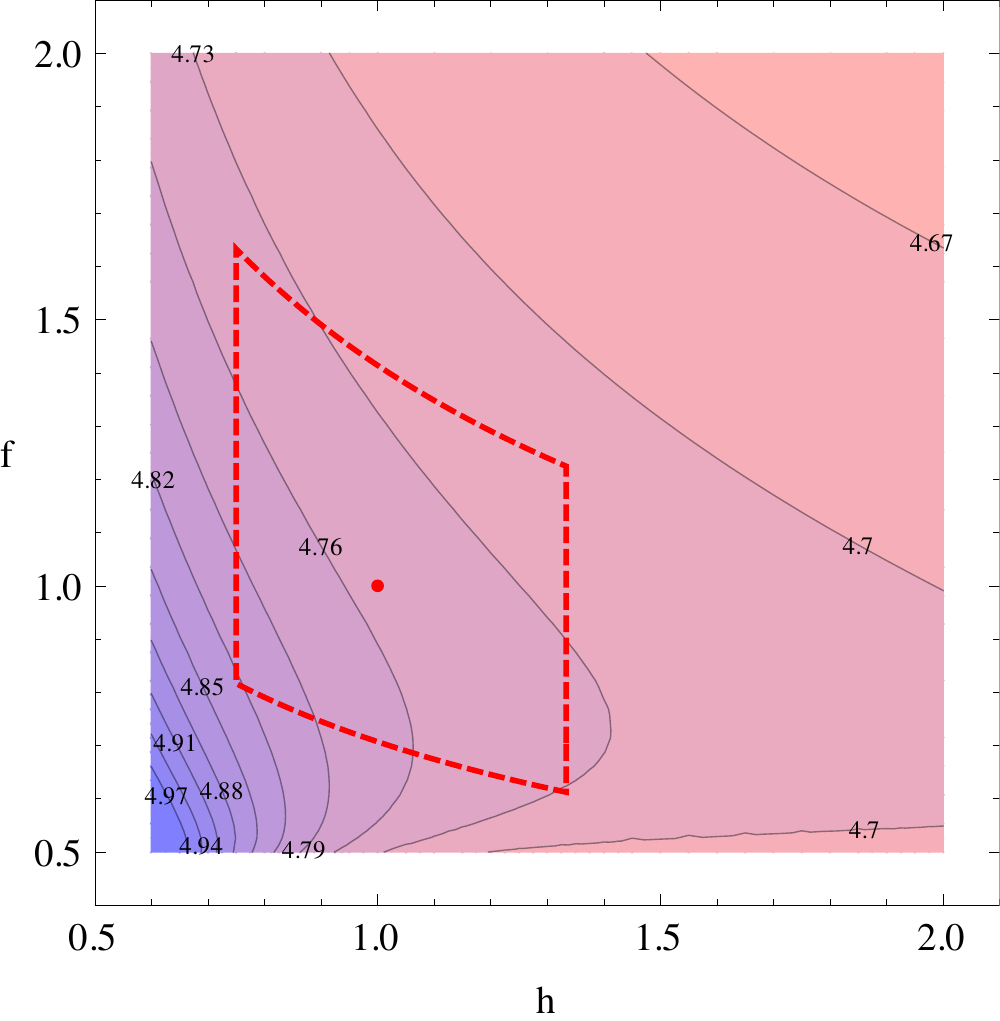}
\end{center}
\caption{Contour plot of the 1S bottom mass determined from $P^{th}_{10}(m_b) =
  P^{exp}_{10}$ as a function of the parameters $h$ and $f$ as defined in the
  text. The different contours are labeled by the respective mass value in GeV. 
The region in the $h$-$f$ plane bounded by the red dashed line represents
the parameter space we scan to determine the variation of the mass, which
contributes to our perturbative error  estimate. The region is defined by $0.75 \le h \le
1/0.75$ and demanding that $0.5\, \mu_U^* \le \mu_U \le 2\, \mu_U^*$, where
$\mu_U^*=m_b \nu^{*2}$. 
The red point inside this area indicates our default values $f=h=1$ for the mass
determination. 
\label{ContourPlot}} 
\end{figure}

The contour plot in Fig.~\ref{ContourPlot} shows the results for the 1S bottom
mass obtained for $n=10$ at NNLL order using the expanded expression 
for the theoretical moment in the
$h$-$f$-plane. The default values 
are $h=f=1$ (red dot). For estimating the
perturbative error we vary $h$ and $f$ around their default values applying the
restriction that the ultrasoft renormalization scale remains 
within the range $0.5\, m_b \nu^{*2} \le \mu_U \le 2\, m_b
\nu^{*2}$. Furthermore we fix the range for the 
variation of $h$ to $0.75 \le h \le 1/0.75$. The two conditions define the region
indicated by the red dashed lines, and we note that for both $h$ and $f$ being
small this region restricts the ultrasoft scale from dropping 
below $0.13\,m_b$ for $n=10$.
\begin{figure}[t]
\begin{center}
\includegraphics[width=0.48\textwidth]{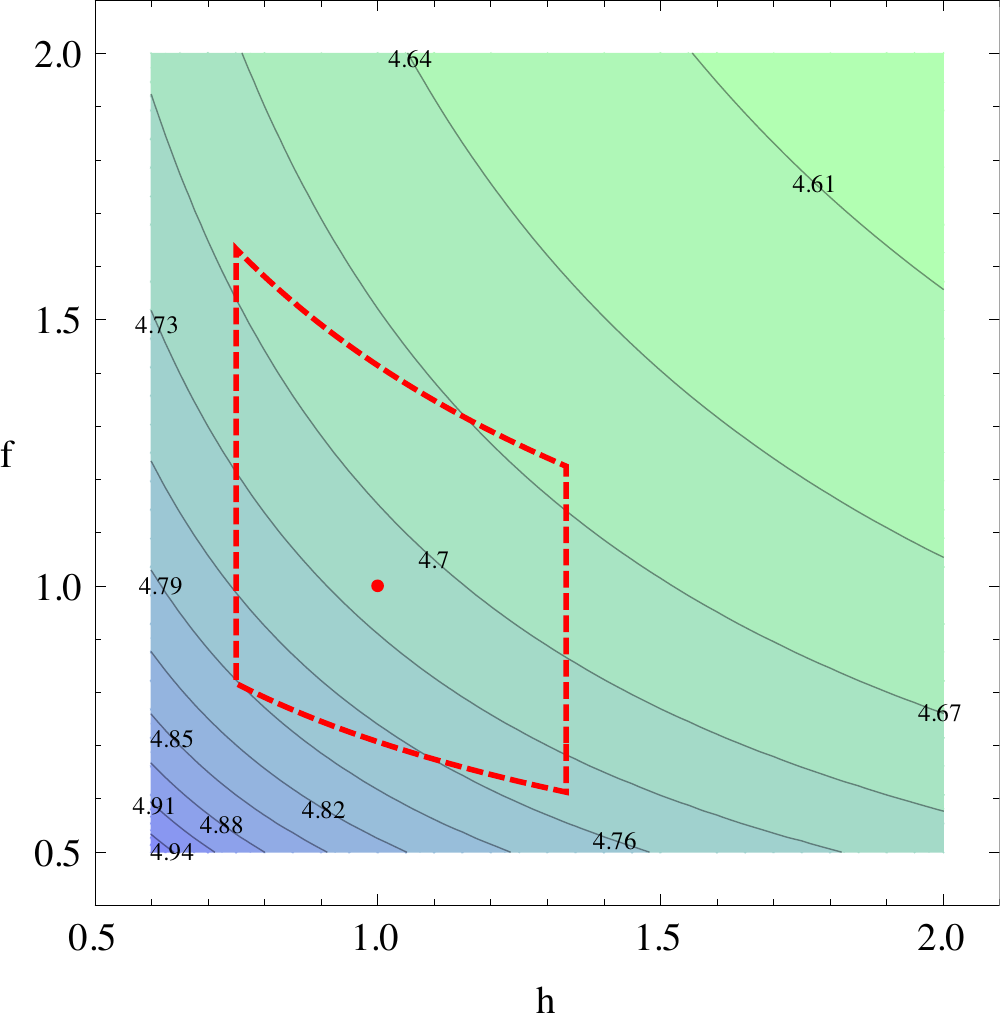}
\quad
\includegraphics[width=0.48\textwidth]{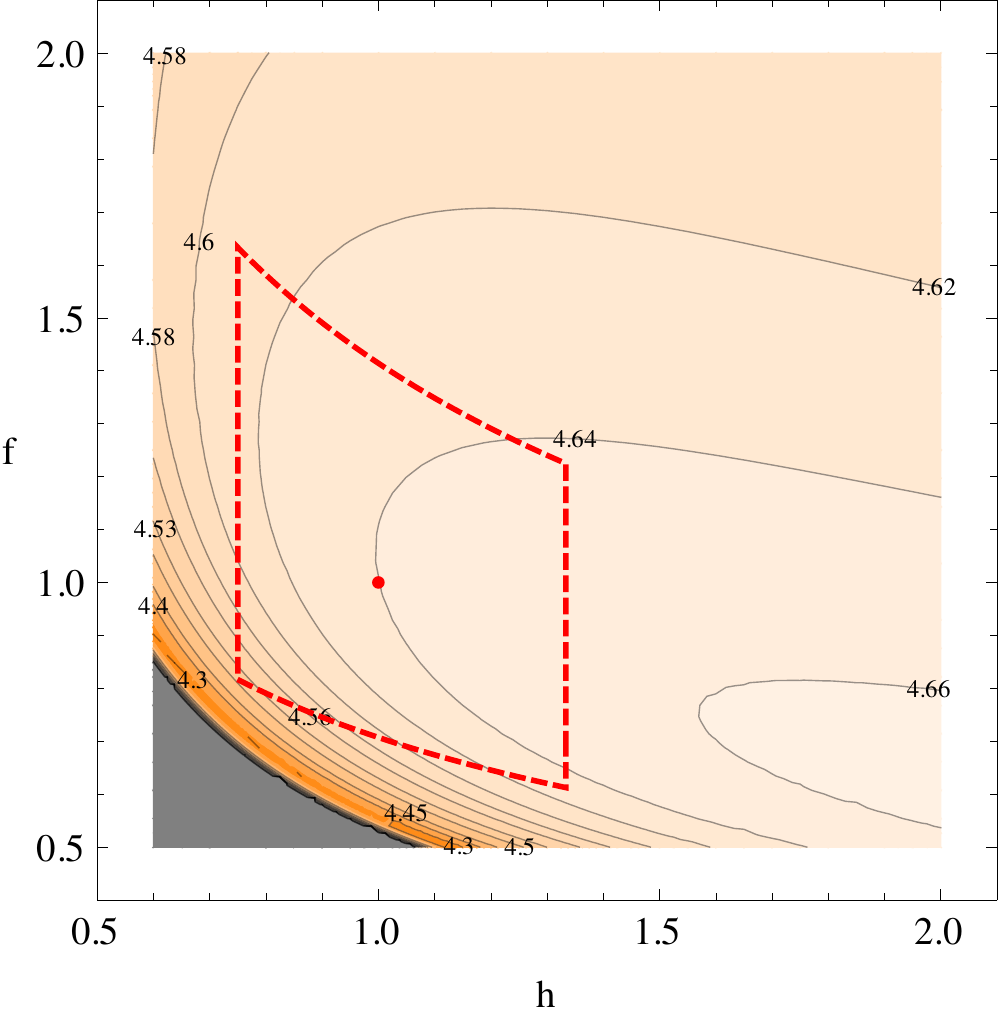}
\end{center}
\caption{Contour plots of the 1S bottom mass contour plots determined from
  the $10$th moment at LL (left panel) and NLL (right panel) order. The gray
  region in the lower left corner of the NLL plot indicates that 
  the equation $P^{th,NLL}_{10}(m_b) = P^{exp}_{10}$ does not have
  a solution related to the ultrasoft scale being too small
  ($\mu_U \lesssim 0.6$~GeV). 
\label{ContourPlot2} }
\end{figure}

Figure~\ref{ContourPlot} displays the overall behavior for the bottom mass
extracted from the NNLL order moments, indicating steeply rising mass values for
decreasing values of $h$ and $f$ corresponding to the matching, soft and ultrasoft scales
being small. However, in the restricted region we use for estimating the
perturbative error, the contours are relatively flat with the default $m_b$
representing a central value within the overall range of values obtained in the
restricted region. 

For comparison we have also displayed in
Fig.~\ref{ContourPlot2} the corresponding bottom mass contour plots obtained from the LL order
 moment (left panel) and from the NLL order moment (right panel). It is a
conspicuous fact that in the restricted $h$-$f$ region the form of the contours
and the range of the bottom masses covered for the LL analysis 
is quite similar to the NNLL order analysis. 
On the other hand, for the NLL order analysis the outcome is quite different,
since the mass values strongly decrease in the small $h$-$f$ region and since the
default bottom mass is very close to the upper end of the obtained range of
bottom mass values. Moreover there is hardly any overlap to the mass range
obtained from the NNLL and LL analyses. 
The overall behavior we see from the
outcome of this analysis at LL, NLL and NNLL order confirms
the slightly marginal character of the perturbative series for the large-$n$
moments we already discussed in Sec.~\ref{sectionpertexp}. 

From our observations it is clear that one
has to conclude that the LL and NLL order renormalization scale variations do
not provide reliable estimations of the perturbative uncertainty at their
respective orders and that the uncertainties are in fact substantially
larger. On the other hand, the 
consistency of the analysis at LL and NNLL order and between the NNLL order bottom mass
results for the expanded and factorized moments, as shown below, are indications
that the NNLL analysis is reliable and that the variations of $f$ and $h$ in the
region described above give a proper estimate of the perturbative NNLL order
uncertainty. We therefore adopt the NNLL order large-$n$ moments and the
procedure as described above to determine the bottom mass and the perturbative
uncertainty for our final numerical analysis following below.

\begin{figure}[t]
\begin{center}
\includegraphics[width=\textwidth]{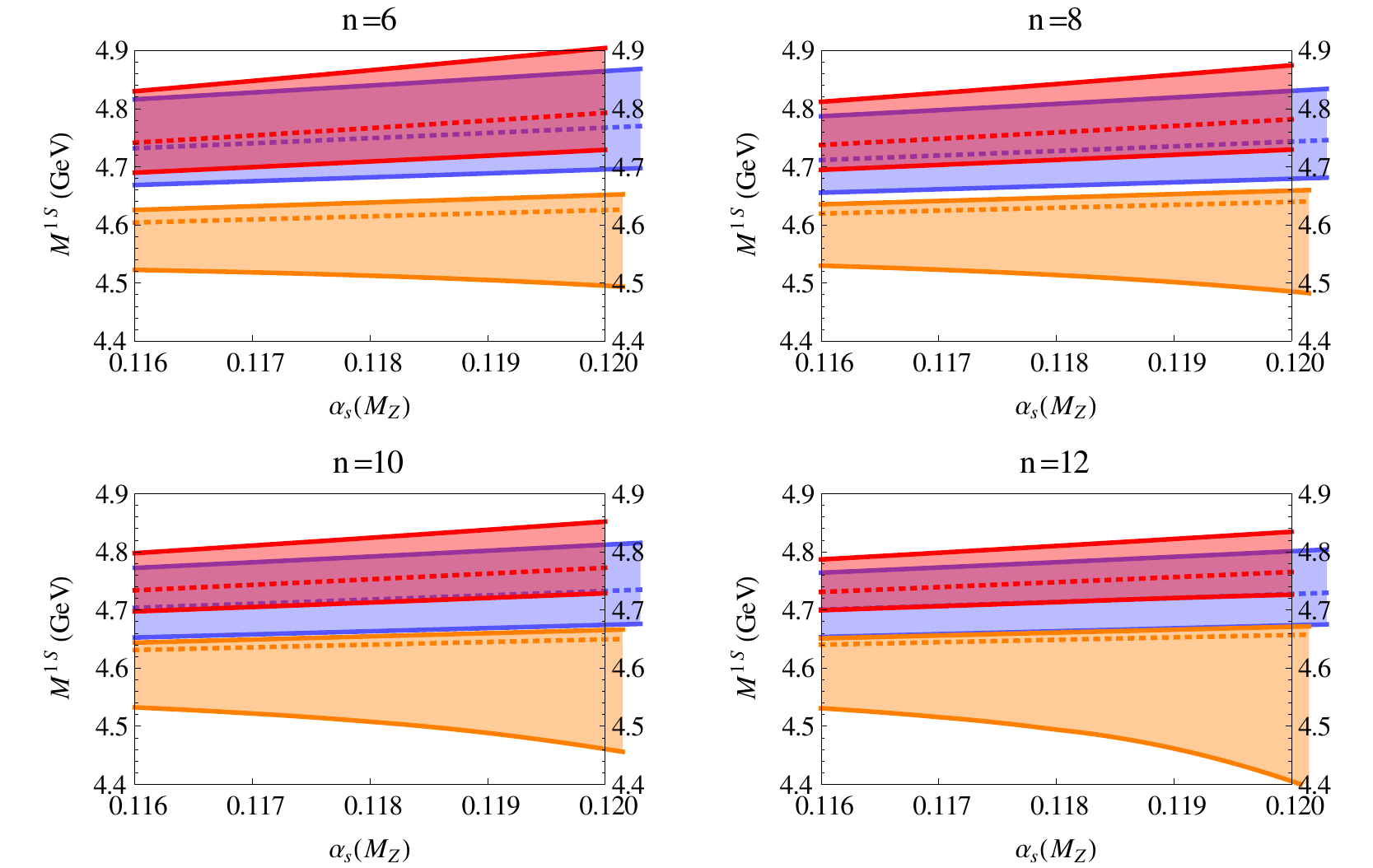}
\put(-249,246){\blau \sf LL}
\put(-255,206){\orange \sf NLL}
\put(-262,264){\rot \sf NNLL}
\end{center}
\caption{1S bottom mass with errors determined from $P^{th}_{n}(m_b) =
  P^{exp}_{n}$ for $n=6,8,10,12$ as a function of $\alpha_s(M_Z)$ using the
  strictly expanded expression for $P^{th}_{n}(m_b)$. As indicated in the panel
  for $n=6$ we plotted for each moment three (partly overlapping) bands with LL
  (blue), NLL (orange) and NNLL (red) precision, respectively. 
The error bands are generated by varying the parameters $f$ and $h$ of the
theoretical moments $P^{th}_{n}$ within the parameter space defined in
Fig.~\ref{ContourPlot} and adding the corresponding experimental error in
quadrature. 
\label{4MomentsExp}} 
\end{figure}
In Fig.~\ref{4MomentsExp} we present the results for the 1S bottom mass from the
moments for $n=6,8,10,12$ in the expanded form as a function of
$\alpha_s(M_Z)$ at NNLL order (red) and for comparison also at LL (blue) and NLL
order (orange). Given the discussion above we will, however, only discuss the
NNLL order results. The respective dashed lines represent the mass for the default
scale choices with $f=h=1$ and the bands come from varying $f$ and $h$ in the
region as discussed in detail above. The masses obtained from the
default scale choices are remarkably consistent for the different $n$ values
exhibiting deviations of less than $20$~MeV  for 
$\alpha_s(M_Z)$ in the region around $0.118$, where the current world average
$\alpha_s(M_Z) = 0.1183 \pm 0.0010$~\cite{Bethke:2011tr},
is located. 
The width of the band from the scale variations, on the
other hand, slightly decreases for increasing $n$ indicating that the
dependence of the moment normalization on the exponent of the mass slightly 
overcompensates the increase of higher order $\alpha_s$ corrections due to the
smaller renormalization scales. 
Overall, the 1S mass slightly increases with $\alpha_s$, but the dependence is
rather weak and linear to a very good approximation so that the error in
$\alpha_s(M_Z)$ only has a rather small impact on the final uncertainty in the
bottom mass. We note that the uncertainties in the bottom mass coming from the
error of the experimental moments amount to 
$(6.5,4.3,3.3,2.7)$~MeV for $n=(6,8,10,12)$ and
are much smaller than the perturbative error
and the uncertainty coming from $\alpha_s$.  
For quoting the final result the experimental uncertainty does not play any
essential role.

\begin{figure}[t]
\begin{center}
\includegraphics[width=\textwidth]{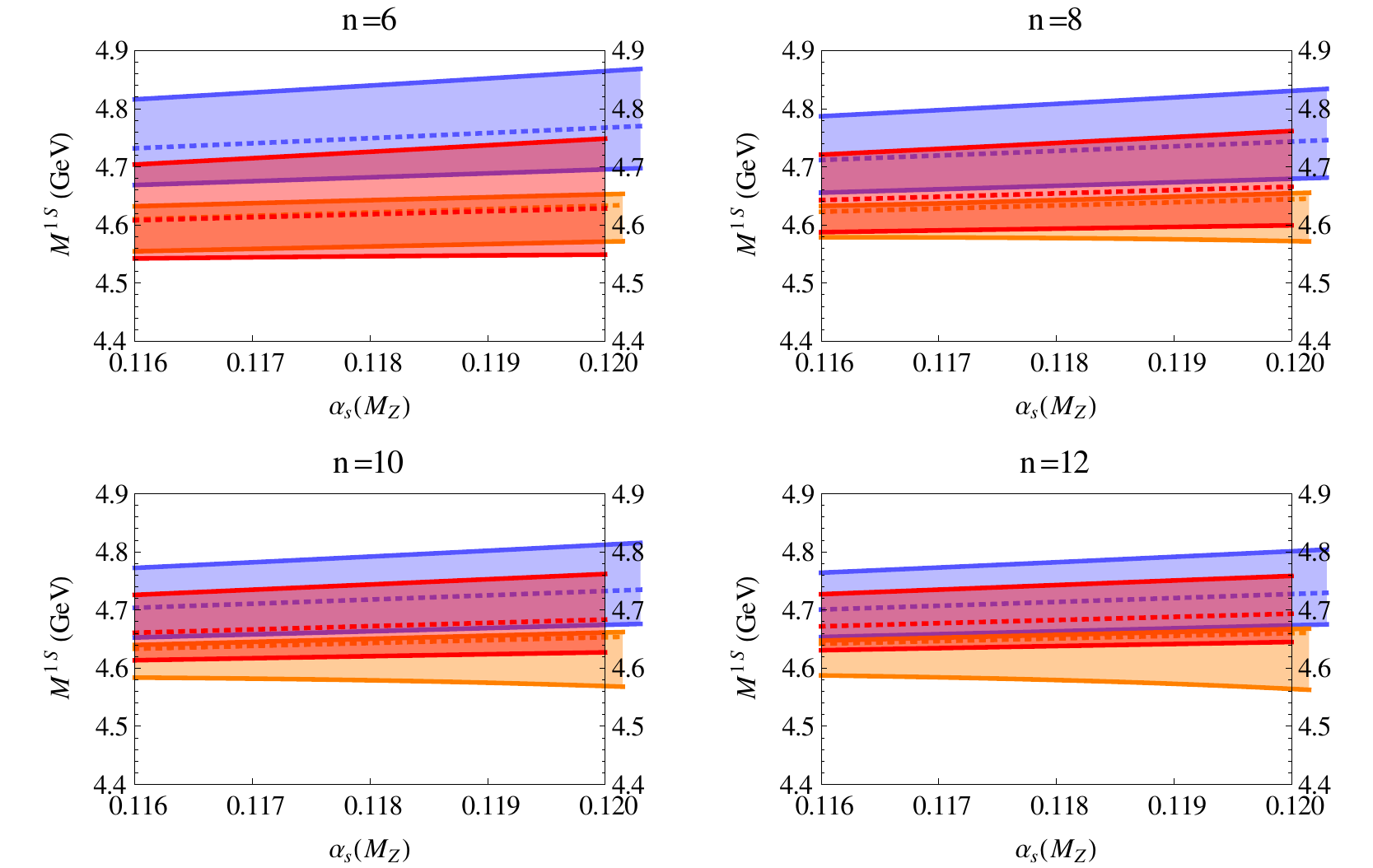}
\put(-249,264){\blau \sf LL}
\put(-255,224){\orange \sf NLL}
\put(-261,243){\rot \sf NNLL}
\end{center}
\caption{Band plots analogous to Fig.~\ref{4MomentsExp} but with the factorized
  expression (as defined in the text) instead of the expanded expression used
  for the theoretical moments. 
\label{4MomentsUnExp}} 
\end{figure}
In Fig.~\ref{4MomentsUnExp} we show the corresponding results for the 1S
bottom mass using the factorized expression for the theoretical moments. 
Here the LL results are identical to the previous analysis based on the expanded
moments. While the width of the NNLL order bands from scale variations are almost
identical to the expanded moment analysis, we see that the NNLL order results
for the mass values are somewhat 
smaller, but consistent within errors with the results for the expanded
moments. The $n$-dependence of the mass values for $\alpha_s$ values around the
world average is about three times larger than for the fully expanded
moments. We note again, that the factorized moments have an inconsistent 
renormalization scaling behavior, and we therefore adopt the moments in the 
expanded form for the main numerical analysis. The consistency of the
results from expanded and the factorized moments, on the other hand, indicates
that the two results are in agreement within their perturbative
uncertainties. 

For quoting our final number for the 1S bottom quark mass we take the default
value obtained from the $10$-th NNLL order moment in expanded form as
the central value and adopt half the size of the band from scale variation as the
perturbative uncertainty. Using the current world average for the strong
coupling 
$\alpha_s(M_Z) = 0.1183 \pm 0.0010$~\cite{Bethke:2011tr} 
we obtain
\begin{align}
M_b^{1S} \, = \, 4.755\, \pm \, 0.057_{\rm pert}\, \pm\, 0.009_{\alpha_s}\,
\pm \, 0.003_{\rm exp}
\,\,\mbox{GeV}
\,.
\label{mbfinal1}
\end{align}
It is worth to emphasize that the $\alpha_s$ error increases by $9.3$~MeV for
each unit of $0.001$ the
uncertainty in $\alpha_s(M_Z)$ is increased.
It is also interesting to quote the result as a function of $\alpha_s(M_Z)$.
Since the dependence of the bottom mass on the value of the strong coupling is
linear to a very good approximation,
we can quote the $\alpha_s$-dependence of the bottom 1S mass by the
following numerical fit formula:  
\begin{align}
M_b^{1S} \, = \, \Big[ 4.755 + \frac{\alpha_s(M_Z)-0.1183}{0.107}\Big]\, 
\pm \, \Big[ 0.057 +\frac{\alpha_s(M_Z)-0.1183}{0.37}\Big]_{\rm pert}\,\,\mbox{GeV} 
\,,
\label{mbfinal2}
\end{align}
where we have dropped the experimental error.
The deviation of the linear approximation in Eq.~(\ref{mbfinal2}) to the exact
result is less than 0.6 MeV for $0.113\le\alpha_s(M_Z)\le 0.120$. 

It is straightforward to convert our result for the 1S mass to the $\msb$
mass by combining the pole-1S mass relation in Eq.~\eqref{mpoletom1s} 
and the pole-$\msb$ mass relation.
For this conversion one has to employ the
$\epsilon$-expansion~\cite{Hoang:1998hm,Hoang:1998ng,Hoang:1999ye} to
consistently cancel the renormalon contributions. 
The systematic cancellation of
the renormalon contributions also requires that, within the conversion, the corrections in the pole-1S mass relation are evaluated at the same renormalization
scale as those in the pole-$\msb$ mass series. 
This can be achieved with the best
convergence using the hard scale $M_b^{1S}$ and requires to use the fixed order
expansion for the pole-1S mass relation which is obtained from
Eq.~(\ref{mpoletom1s}) for $\nu=1$ and $h=1$. 
Since the conversion formula has a significant $\alpha_s$-dependence due
to a term $\propto\alpha_s m_b$ arising in the series relating the $\msb$ and pole
masses one needs to convert using the full $\alpha_s$-dependence of the 1S mass
result as displayed in Eq.~(\ref{mbfinal2}). 

Using again the world average for the strong coupling as an input we
then obtain
\begin{align}
\overline m_b(\overline m_b) \, = \, 4.235\, \pm \, 0.055_{\rm pert}\,
\pm \, 0.003_{\rm exp}
\,\,\mbox{GeV}
\,,
\label{msbfinal1}
\end{align}
where we have added an additional conversion error of $15$~MeV to the perturbative
uncertainty coming from half the size of the three-loop correction\footnote{
This corresponds to the ${\cal O}(\epsilon^3)$ terms in the epsilon expansion,
see e.g. Eq.~(99) of Ref.~\cite{Hoang:2000fm}.} in the perturbative series. The resulting value for the perturbative
uncertainty is, however, still smaller than for the 1S mass due to the
conversion series. The uncertainty coming from the value of
$\alpha_s$ reduces to less than $0.5$~MeV, as the sizable $\alpha_s$ dependence
in the 1S-$\msb$ conversion series mentioned above is anticorrelated to the
monotonically increasing $\alpha_s$ dependence of the 1S mass. Remarkably these
two effects cancel almost entirely. We have therefore
dropped the $\alpha_s$ induced error from Eq.~(\ref{msbfinal1}). In analogy to
Eq.~(\ref{mbfinal2}) we can also give the result for the $\msb$ mass showing the
full $\alpha_s$ dependence as a linear fit function:
\begin{align}
\overline m_b(\overline m_b) = \, 4.235 \, 
\pm \, \Big[ 0.055 +\frac{\alpha_s(M_Z)-0.1183}{0.41}\Big]_{\rm pert}\,\,\mbox{GeV} 
\,,
\label{msbfinal2}
\end{align}
where we did not include a linear $\alpha_s$ fit term for the central value as
it contributes less than one MeV  for $0.113\le\alpha_s(M_Z)\le 0.120$. Again
we have dropped the experimental error. 

Our result for the bottom $\msb$ mass $\overline m_b(\overline m_b)$ should be
compared with the corresponding result obtained in the analysis of
Ref.~\cite{Pineda:2006gx}. Their analysis was also based on RGI large-$n$
moments having NNLL order input, but was using only the NLL order result for the
anomalous dimension of the leading order current Wilson coefficient $c_1$.
Thus their theoretical moments were missing the numerically important NNLL
contribution to the evolution of $c_1$ shown in Fig.~\ref{c1Plot}. Since $c_1$
enters as a square, their theoretical moments were by about $30$\% lower than
ours. For a single fit for the $10$-th moment one therefore expects, due to the
$1/m_b^{2n}$ overall dependence on the mass as shown in Eq.~(\ref{Pnth}), that
their mass result is by about $1.3$\% lower than ours. This amounts to about
$50$~MeV and is consistent with the result $\overline m_b(\overline m_b)=4.19\pm
0.06$~MeV obtained in their analysis. The perturbative uncertainty estimated in
their work was quoted between $45$ and $65$~MeV and is numerically similar to
ours. We note, however, that they determined the uncertainty from the difference
in the bottom mass obtained from the NLL order and their incomplete NNLL order
moments. 

We also note that our result for the $\msb$ mass is not quite compatible with
the results given in Ref.~\cite{Chetyrkin:2009fv} 
($\overline m_b(\overline m_b) =4.163\pm 16$~MeV) or even in
Ref.~\cite{Bodenstein:2011fv}
($\overline m_b(\overline m_b) =4.171\pm 9$~MeV) which were obtained from ${\cal
  O}(\alpha_s^3)$ fixed order analyses related to low-$n$ $\Upsilon$ sum rules
and taking their uncertainties exactly as quoted.
Here we note again that we have treated the charm as a massless quark. In previous
fixed-order analyses~\cite{Hoang:1999us,Hoang:2000fm} it 
was found that the non-zero charm mass causes shifts in the bottom quark mass
obtained from large-$n$ sum rules between $-20$ and $-30$~MeV. This sizable
corrections is related to the fact that the charm mass is similar to the inverse
Bohr radius $\alpha_s m_b$ and therefore not $1/m_b$-suppressed.  
At this time the renormalization group improved non-relativistic effective
theory approach has not yet been extended systematically to account for massive
virtual quark thresholds. The required conceptual developments are theoretically
interesting and might contribute reconciling the discrepancy. 
For comparing our result to bottom mass determinations from other physical
quantities we refer to Ref.~\cite{Beringer:1900zz}.

\section{Fixed-Order Moments and Multiple Moment Fits}
\label{sectiondiscussion}

Before RGI predictions based on the extended versions of NRQCD were possible a
number of fixed-order analyses were carried out at the NNLO 
level~\cite{Melnikov:1998ug,Hoang:1998uv,Beneke:1999fe,Hoang:1999ye,Hoang:2000fm}.  
As mentioned in Sec.~\ref{sectionintro}, the fixed-order moments showed a rather
bad perturbative behavior, and the different analyses addressed
this issue in different ways in order to achieve bottom mass determinations with
small uncertainties. We can turn our RGI moments into fixed-order moments by
setting the matching, soft and ultrasoft scales equal to the soft scale $\mu_S$,
except for the current Wilson coefficients, where we keep the matching scale
$\mu_h$. This leads to a logarithm $\ln(\mu_S/\mu_h)=\ln(f\nu^*)$ appearing in
the current Wilson 
coefficient $c_1$ whose coefficient does not depend on the regularization 
scheme nor the conventions used for the potentials.
Although the form of the resulting moments do not agree exactly with either one of the analyses in
Refs.~\cite{Melnikov:1998ug,Hoang:1998uv,Beneke:1999fe,Hoang:1999ye,Hoang:2000fm} due to
differences in the regularization scheme, the expansion scheme or in the
conventions for the potentials, the resulting fixed-order moments provide a good
insight into the improvement obtained by the renormalization group logarithms
summed into the RGI moments.

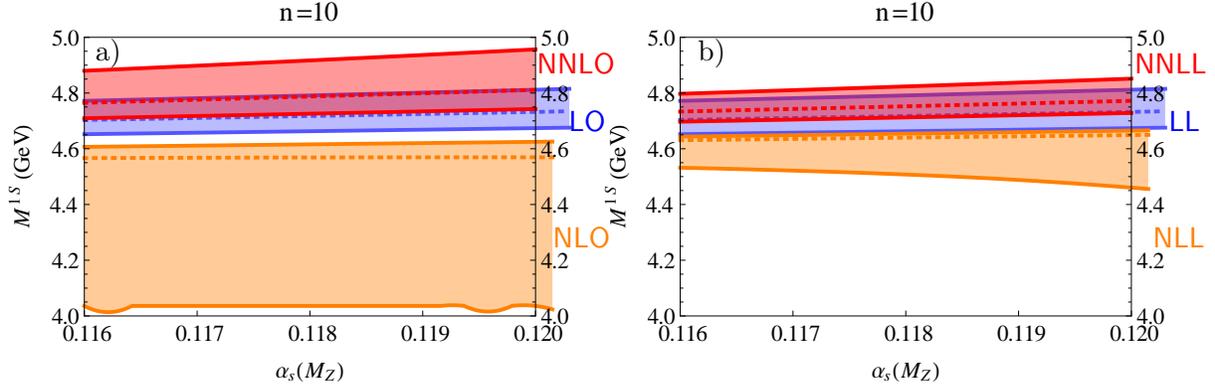
\begin{figure}[t]
\includegraphics[width=\textwidth]{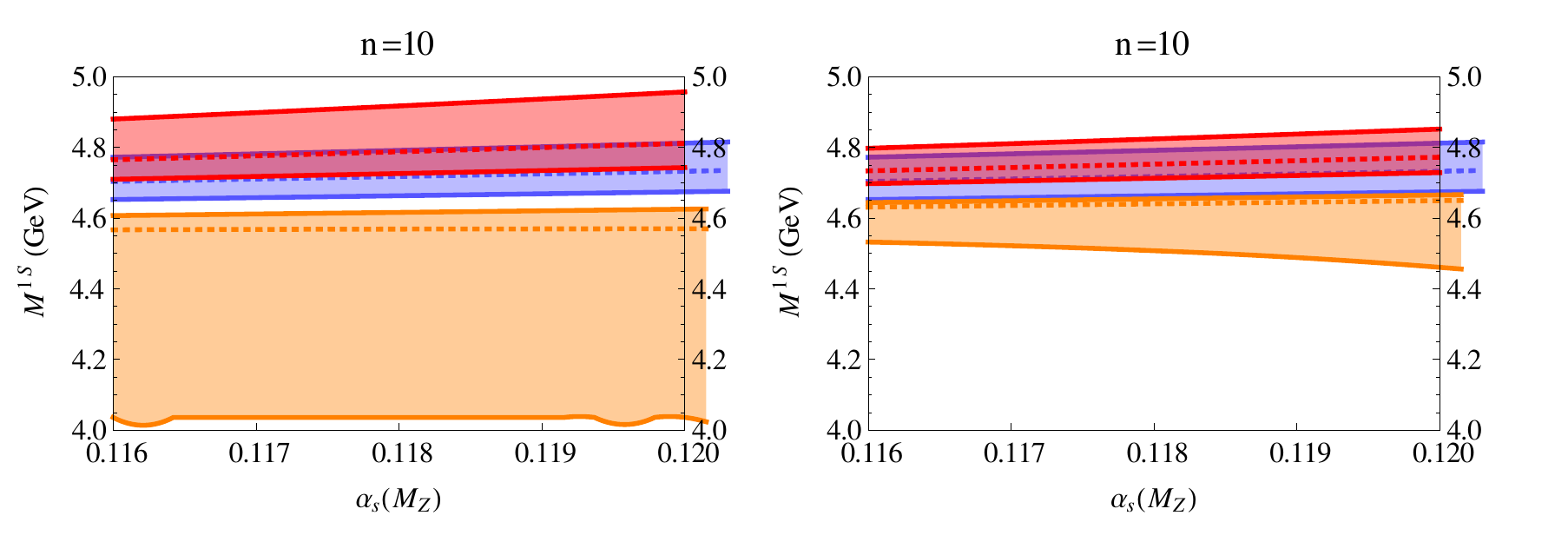}
\put(-251,103){\blau \sf LO}
\put(-257,59){\orange \sf NLO}
\put(-263,125){\rot \sf NNLO}
\put(-24,103){\blau \sf LL}
\put(-30,59){\orange \sf NLL}
\put(-37,125){\rot \sf NNLL}
\put(-430,130){a)}
\put(-202,130){b)}
\caption{Comparison of the masses obtained from the fixed order (a) and RGI
  calculation (b) of the 10-th moment, $P^{th}_{10}(m_b) = P^{exp}_{10}$. In
  panels a and b we show the  mass values with LO, NLO, NNLO and LL, NLL, NNLL
  accuracy, respectively. The corresponding error bands were generated in the
  same way as in Fig.~\ref{4MomentsExp}.
  Concerning panel a, we note that for some low $m_b$ values in the NLO band and
  the associated values for $h$ and $f$ the ultrasoft coupling $\alpha_S(\mu_U)$ 
  reaches $0.65$ causing numerical instabilities.
\label{RGIvsFixedOrder}} 
\end{figure}
The 1S bottom mass results obtained by carrying out a single moment
fit for our fixed-order moment
for $n=10$ in the expanded form and using the same variations of $h$ and
$f$ we employed in the RGI analysis,\footnote{The region in the
  $h-f$ plane we use in our analysis leads to the variations
  $0.75 m_b\le\mu_h\le 1.33 m_b$ and $0.32 m_b\le \mu_S \le 0.84 m_b$
  for the matching and soft scales, respectively. For the matching scale the
  variation is 
  somewhat smaller than what was employed in the fixed-order analysis of
  Refs.~\cite{Melnikov:1998ug,Hoang:1998uv,Beneke:1999fe,Hoang:1999ye,Hoang:2000fm} but
  roughly agrees with those analyses for the soft scale.}
 which now only cause correlated variations
in the matching and the soft renormalization scale,
is shown in Fig.~\ref{RGIvsFixedOrder}.
The result should be
compared with the outcome using the RGI NNLL order moment $P_{10}^{th}$ shown in the
lower left panel of Fig.~\ref{4MomentsExp}, which we have for 
convenience displayed once more in the right panel of
Fig.~\ref{RGIvsFixedOrder}. Because at LL 
order the soft scale is the only renormalization scale that arises in the
theoretical moment, the LL RGI and LO fixed-order results agree
exactly. Substantial differences are, however, visible for the NLL vs.\ NLO 
results where the fixed order result shows even lower bottom mass values and a
substantially larger scale variation. Comparing the NNLL and NNLO results we see
that the fixed-order results cover larger bottom masses and exhibit scale
variations that are twice as big as for the RGI results. We clearly see the
improvement related to summing the logarithmic contributions within the RGI
approach. Determining the final
result for the 1S bottom mass from the NNLO moment in the same way as we did it
in our NNLL RGI analysis we would obtain for $\alpha_s(M_Z)=0.1183\pm 0.001$ the
result
$M_b^{1S}=4.791\pm0.097_{\rm pert}\pm 0.012_{\alpha_s}\pm  0.003_{\rm exp}$~GeV 
(and thus 
$\overline m_b(\overline m_b)=4.269\pm0.090_{\rm pert}\pm 0.002_{\alpha_s}
\pm 0.003_{\rm exp}$~GeV
for the $\msb$ mass), 
which is compatible with our results from Eqs.~(\ref{mbfinal1}) and
(\ref{msbfinal1}), but with a substantially larger perturbative uncertainty. 
Our fixed-order result is in agreement with Ref.~\cite{Beneke:1999fe} 
($\overline m_b(\overline m_b)=4.26\pm 0.09_{\rm pert}\pm 0.01_{\alpha_s}
\pm 0.02_{\rm exp}$~GeV) which also used a NNLO fixed-order single moment fit
for $n=10$ and at the time of the analysis accounted for larger strong coupling
and experimental errors.

\begin{figure}[t]
\includegraphics[width=\textwidth]{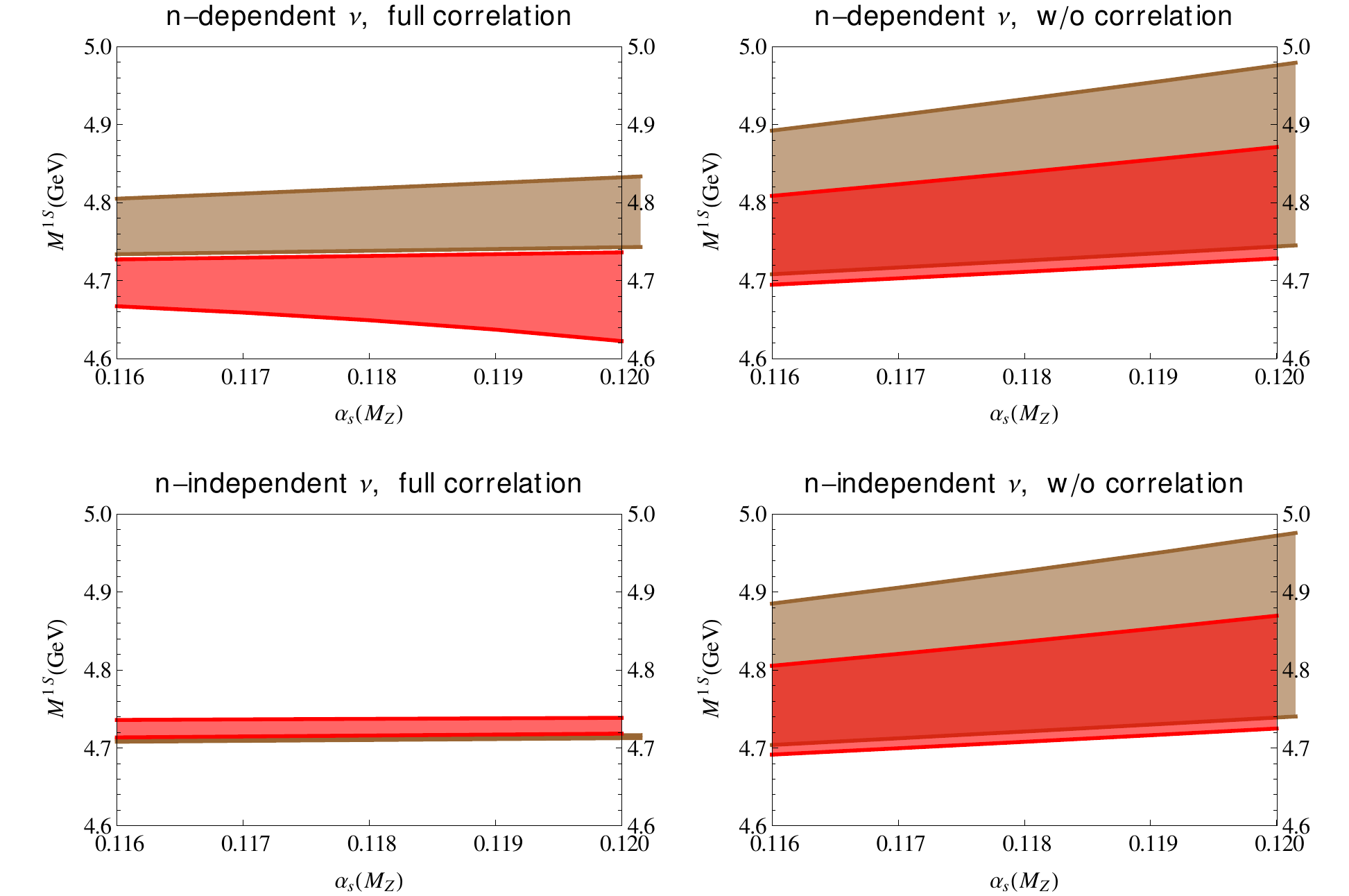}
\put(-251,200){\rot \sf NNLL}
\put(-255,252){\color{brown} \sf NNLO}
\caption{Multiple moment fits for the 1S mass with $n=\{4,6,8,10\}$. Each panel
  shows two bands associated with a fit using the NNLO fixed order
  moments (brown) and the NNLL RGI moments (red), respectively. The
  bands are generated by scanning the usual $h-f$ region (see
  Fig.~\ref{ContourPlot}). Fits scanning an $n$-dependent $\nu$ range with
  $\nu^*=(1/\sqrt{n}+0.2)$ and an $n$-independent $\nu$ range with
  $\nu^*=(1/\sqrt{7}+0.2)$ while allowing for full/no correlation of the
  experimental moments are shown as indicated by the labels of the individual
  panels. 
\label{MultipleMomentFitsexp}
}
\end{figure}

Another instructive analysis is related to carrying out simultaneous fits to
several moments using a $\chi^2$ analysis. It is straightforward to carry out
such fits with our moments as well. In Fig.~\ref{MultipleMomentFitsexp} we
show the allowed bottom mass ranges obtained from $\chi^2$-fits based
on the moments for $n={4,6,8,10}$ where brown refers to the fixed-order NNLO
analysis and, for comparison, red to the RGI NNLL order analysis. The upper
panels show the outcome for the $n$-dependent assignment of the renormalization
scales and the variation in the $h-f$ plane as described for the previous
analyses, i.e.\ the choice of scales for the moments entering the $\chi^2$
function is $n$-dependent. Here the left panel is obtained using the full
experimental covariance matrix (based on treating the experimental
input data for masses, widths and $R$-ratio values as 
statistically independent), and
the right panel is  obtained treating the moments as statistically independent.
The result for independent moments 
agrees almost exactly with the result of the $n=10$ single moment fit, which is
easily understood since $P_{10}^{exp}$ has the smallest error and thus the highest weight
for uncorrelated moments.

Accounting for the correlation, on the other hand,
leads to a different outcome, where the NNLL order bottom mass values are lower
and the NNLO fixed order results lead to much smaller scale variations
 compared to the single moment or the uncorrelated fits. We have
traced this behaviour back to the fact that the correlation coefficients are
very close to unity and that the $\chi^2$ function fits the $n$-dependent shape
of the moments rather than the actual values of the moments. 
Together with the correlated scale dependence of the theory moments this leads to
the observed strong cancellation of the scale variation.
This is also
associated to rather large minimal $\chi^2/\rm dof$ values with an average of around
$50$. From this we conclude that the correlations are too large and the
resulting scale variations are unreliable. 
Accounting in addition for the fact that the theory uncertainties are much bigger than the experimental ones,
we conclude that single moment fits seem to be the 
only practical option to extract the bottom mass.

This conclusion is also reconfirmed by the analysis shown in the lower
two panels. Here we have displayed the results for the bottom mass using the
same $\chi^2$ fit method as for the respective upper two panels, but now using 
the renormalization scales adopted for $n=7$ (the average of the $n$ 
values used for the fit) for all moments. 
Treating the experimental moments as uncorrelated
(right lower panel) the results for the bottom mass are very similar as for the
moments with $n$-dependent scales (shown in the right upper panel) since again
the $10$-th moment has the highest weight in the fit.  On the
other hand, including the correlation we find that almost all scale variations
cancel. For the NNLL moments the resulting scale variation is between $25$ and
$30$~MeV, while for the NNLO moments it is below $5$~MeV. This remarkable
cancellation can be traced back to the fact that the 
experimental correlation matrix puts the highest weight to a linear combination
of moments that minimized the sensitivity to the uncertainties of the $\Upsilon$
electronic widths, which represent the biggest source of uncertainty for the
large-$n$ experimental moments. Using the same renormalization scale values for
all moments reduces very effectively the sensitivity to the electronic width of
the perturbative $\Upsilon$ bound states entering the theoretical moments. Since
the theoretical electronic widths, and the theoretical $R$-ratio in general, 
are the biggest source of scale variation in the theory moments, a common
choice of renormalization scale in all moments, leads to the observed reduction
of scale variation. 

Given our observation that the experimental correlation of
the moments is too strong to allow for reliable multiple moment fits, we have to
conclude that the tiny scale variation shown in the lower left panel is an
artifact of the scale choice and does not represent an estimate of the
perturbative uncertainty.
Overall the unreliable multiple moment fits are a consequence of the fact that the theory errors are 
significantly larger than the experimental ones.
We note that fixed-order multiple moment fits with
$n$-independent renormalization scales based on several different sets of
moments were used in Refs.~\cite{Hoang:1998uv,Hoang:1999ye,Hoang:2000fm}. The theoretical
uncertainties quoted in these analyses originated mainly from using the
different sets of moments and not from the scale variation.

\section{Conclusion}
\label{sectionconclusion}

In this work we have determined the 1S bottom quark masses from
large-$n$ $\Upsilon$ sum rules with renormalization group improvement. Since for
large values of $n$ the quark dynamics entering the moments is non-relativistic,
the experimental moments are dominated by the well known data on the $\Upsilon$
resonances and the experimentally unmeasured continuum contributions are
not relevant. The theoretical computation of the moments involves a simultaneous
expansion in $\alpha_s$ and $1/\sqrt{n}$ and is carried out within the
framework of non-relativistic effective field theories. In general, within the
non-relativistic framework where $n$ is large, one
has to treat terms $\propto\alpha_s/\sqrt{n}$ and $\propto\alpha_s\ln(n)$ of
order unity. For the renormalization group improved treatment both types of
terms are summed to all orders following the generic scheme of
Eq.~(\ref{counting}). We have carried out this scheme using the vNRQCD
(``velocity'' NRQCD) 
formalism originally devised in Ref.~\cite{Luke:1999kz} and
extended in a number of subsequent publications as discussed in the main body of
this paper. We have also determined the $\msb$ bottom quark mass $\overline
m_b(\overline m_b)$. We note that the conversion involves a sizable dependence
on the value of the strong coupling $\alpha_s$. While we find that the bottom 1S
mass slightly increases with $\alpha_s$, the  $\msb$ mass $\overline
m_b(\overline m_b)$ turns out to be essentially $\alpha_s$-independent.

In this work we have for the first time carried out a renormalization group improved
analysis using all available NNLL order results. At the NNLL order level
currently all corrections are known except for the so-called soft mixing
corrections to the anomalous dimension of the leading order heavy quark pair
production current. We have given arguments showing that the known full set of
NNLL order ultrasoft corrections numerically dominate over the soft
contributions and that the uncertainty coming from the set of unknown soft
corrections are negligible compared to the large ultrasoft contributions and
their remaining scale variation at the NNLL order level. Our results can
therefore be considered as the outcome of a full NNLL order analysis.

Prior to our work a number of large-$n$ analyses were carried out, which
were based on the so-called fixed-order approach, where only the terms
$\propto\alpha_s/\sqrt{n}$ were summed up
systematically~\cite{Melnikov:1998ug,Hoang:1998uv,Beneke:1999fe,Hoang:1999ye}.  
One important characteristic of these fixed-order analyses was that they
exhibited rather bad convergence properties. We found that the summation of the
logarithms contained in our moments improves the convergence
considerably. Nevertheless the perturbative series for the renormalization group
improved moments shows sizable corrections which made a careful examination of
the perturbative uncertainty of our NNLL order prediction mandatory. 

Our final results have been
presented in Eqs.~(\ref{mbfinal1}--\ref{msbfinal2}) of
Sec.~\ref{sectionanalysis}.

\acknowledgments{
The work of P.~R. is partially supported by MEC (Spain) under grants FPA2007-60323 and FPA2011-23778 and by the
Spanish Consolider-Ingenio 2010 Programme CPAN (CSD2007-00042). 
We thank Vicent Mateu and Bahman Dehnadi for providing us with the code for the
experimental moments and Vicent Mateu for discussion. 
}
 

\appendix
\begin{section}{NNLL calculation of the moments}
\label{Apprhos}

The calculation of the contributions $\varrho_{n,1}$ and $\varrho_{n,2}$ to the
$n$-th NNLL moment in Eq.~(\ref{Pnth}) follows the lines of
Ref.~\cite{Hoang:1998uv}. 
The results can be written as 
\begin{align}
\varrho_{n,1} = &\;
[\varrho_{n,1}]_c + [\varrho_{n,1}]_{\rm kin} + 
\left( {\cal V}_2^{(s)}(\nu) + 2 {\cal V}_s^{(s)}(\nu) \right)\,[\varrho_{n,1}]_{\delta}
      + {\cal V}_r^{(s)}(\nu) \,[\varrho_{n,1}]_{r}
\nn\\[1.5 ex] & 
      - C_A C_F \as^2 \,[\varrho_{n,1}]_{\rm CACF}
      + \frac{C_F^2}{2} \as \,[\varrho_{n,1}]_{\rm CF2}
\nn\\[1.5 ex] & 
      + \as^2 \,{\cal V}_{k1}^{(s)}(\nu) \,[\varrho_{n,1}]_{k1}
      + \as^2 \, {\cal V}_{k2}^{(s)}(\nu) \,[\varrho_{n,1}]_{k2}
\label{rho1}
\end{align}
and
\begin{align}
\varrho_{n,2} = & \,
\frac{3}{4n}\,\bigg[\,
2 +\frac{8}{3}\,\sqrt{\pi}\,\phi - 
4\,\sqrt{\pi}\,\sum\limits_{p=2}^{\infty}\,\phi^p\,
\frac{2\,(p-3)}{3}\,
\frac{\zeta_p}{\Gamma(\frac{p-1}{2})}
\,\bigg]
\,.
\label{rho2}
\end{align}
We abbreviate the strong coupling at the matching, soft and ultrasoft scales
as $\ah=\alpha_s(\mu_h)$,
$\as=\alpha_s(\mu_h f \nu^*)$ and $\au=\alpha_s(\mu_h f^2{\nu^*}^2)$,
respectively, 
and define
\begin{align}
\phi & \equiv \, \frac{a \, \sqrt{n}}{2} 
\quad , \quad a \equiv - \frac{1}{4\pi}\, {\cal V}_{c,\rm eff}^{(s)}(\nu) 
\,.
\label{phi}
\end{align}
The expressions for the potential Wilson coefficients, ${\cal V}_X^{(s)}$, can
be found in Ref.~\cite{Hoang:2002yy}. 
Except for the contribution of the Coulomb potential $[\varrho_{n,1}]_c$, which
contributes at LL and NLL order, all the other terms contained in 
Eqs.~(\ref{rho1},\ref{rho2}) constitute NNLL order contributions. The LL
Coulomb contribution reads 
\begin{align}
[\varrho_{n,1}]_c^{\rm LL}  & = \,
1 + 2\,\sqrt{\pi}\,\phi + 
4\,\sqrt{\pi}\,\sum\limits_{p=2}^{\infty}\,\phi^p\,
\frac{\zeta_p}{\Gamma(\frac{p-1}{2})}
\,.
\label{rhocLL}
\end{align}
and differs from the fixed order LO result given in Ref.~\cite{Hoang:1998uv}
only through the definition of the variable $\phi$ which in the fixed-order
treatment involves the replacement $a\to C_F\alpha_s(\mu_S)$. The difference
constitutes a NNLL order correction. 
For the analytic expressions of the NLL and NNLL contributions to $[\varrho_{n,1}]_c$ 
coming from the one- and two-loop corrections to the 
Coulomb potential we refer the reader to the expressions (57-63) in
Ref.~\cite{Hoang:1998uv}, where the identification $\mu_{\rm
  soft}=\mu_S$ is understood. 

The contribution to the Coulomb potential arising
from the QED photon exchange is formally a NLL order correction,
when we count the electromagnetic coupling $\alpha_{\rm em}$ as
${\cal O}(\alpha_s^2)$. The correction can be implemented in a straightforward
way with the replacement  
${\cal V}_{c,\rm eff}^{(s)} \to {\cal V}_{c,\rm eff}^{(s)} - 4\pi Q_b^2
\,\alpha_{\rm em}$ 
in Eq.~(\ref{rhocLL}).
However, numerically one has to compare $C_F\alpha_s$ with 
$Q_b\alpha_{\rm em}$, and for the strong coupling at the soft scale one obtains
that $(Q_b\alpha_{\rm em})/(C_F\alpha_S)^2$ is ${\cal O}(1\%)$ which makes QED
effects entirely negligible for our NNLL order analysis. For an implementation of
QED effects a fixed-order treatment is therefore well justified. Apart from
the QED effect in the Coulomb potential one then only has to consider a NNLL
order correction in the matching condition for the current coefficient $c_1$ 
given in Eq.~(\ref{c1match}) through the replacement $C_F\alpha_s\to
C_F\alpha_s+Q_b\alpha_{\rm em}$ in the NLL order correction. We find that this
affects the bottom quark mass by less then $1$~MeV, which confirms the
observation made in Ref.~\cite{Pineda:2006gx}.

The remaining non-Coulombic contributions to $\varrho_{n,1}$ differ from
Ref.~\cite{Hoang:1998uv} because we use the differing conventions for the
potentials from Refs.~\cite{Hoang:2002yy,Hoang:2003ns} and dimensional
regularization as the regularization scheme.
The corresponding results are new and read
\begin{align}
[\varrho_{n,1}]_{\rm kin} = & \,
\frac{1}{n}
\,\bigg[\,-\frac{3}{8} -
        4 \,\sqrt{\pi}\,\sum\limits_{p=2}^{\infty}\,\phi^p\,\frac{(p-1)(p-3)}{8}\,
        \frac{\zeta_p}{\Gamma(\frac{p-1}{2})}
\,\bigg]
\nn\\[1.5 ex] & 
+ \frac{2}{n} \, f_n(\phi) + \frac{a^2}{2} \left( \frac{3}{2}-\ln 2+\ln (\mu_S /m_b)\right) \,[\varrho_{n,1}]_c^{\rm LL}
\,,
\label{rhokin}
\end{align}
\begin{align}
[\varrho_{n,1}]_{\delta}  & = \,
-\frac{1}{4\pi}
\,\bigg[\,
        \frac{8}{a\, n} \, f_n(\phi) -2 a \left( -\frac{1}{2}+\ln 2-\ln (\mu_S /m_b) \right) \,[\varrho_{n,1}]_c^{\rm LL}
\,\bigg]
\,,
\label{rhodelta}
\end{align}
\begin{align}
[\varrho_{n,1}]_{r}  = & \,
-\frac{1}{4\pi}
\,\bigg[\,
        \frac{8}{a\, n} \, f_n(\phi) -2 a \left( -\frac{3}{2}+\ln 2-\ln (\mu_S /m_b) \right) \,[\varrho_{n,1}]_c^{\rm LL}
\,\bigg]
\nn\\[1.5 ex] & 
-\frac{1}{4\pi a \,n}
\,\bigg[\,
        2\sqrt{\pi}\,\phi + 4\,\sqrt{\pi}\,\sum\limits_{p=2}^{\infty}\,\phi^p\,
        \frac{\zeta_p(3-p)p}{2\,\Gamma(\frac{p-1}{2})}
\,\bigg]
\,,
\label{rhoVr}
\end{align}
\begin{align}
[\varrho_{n,1}]_{\rm CACF}  & = \,
-\frac{1}{2 a}
\,\bigg[\,
        \frac{8}{a\, n} \, f_n(\phi) -2 a \left( -\frac{5}{4}+\ln 2-\ln (\mu_S /m_b)\right) \,[\varrho_{n,1}]_c^{\rm LL}
\,\bigg]
\,,
\label{rhoVCACF}
\end{align}
\begin{align}
[\varrho_{n,1}]_{\rm CF2}  & = \,
-\frac{1}{2 a}
\,\bigg[\,
        \frac{8}{a\, n} \, f_n(\phi) -2 a \left( -1+\ln 2-\ln (\mu_S /m_b) \right) \,[\varrho_{n,1}]_c^{\rm LL}
\,\bigg]
\,,
\label{rhoVCF2}
\end{align}
\begin{align}
[\varrho_{n,1}]_{k1}  & = \,
-\frac{3}{a}
\,\bigg[\,
        \frac{8}{a\, n} \, f_n(\phi) -2 a \left( -\frac{17}{12} + \ln 2-\ln (\mu_S /m_b) \right) \,[\varrho_{n,1}]_c^{\rm LL}
\,\bigg]
\,,
\label{rhoVk1}
\end{align}
\begin{align}
[\varrho_{n,1}]_{k2}  & = \,
-\frac{2}{a}
\,\bigg[\,
        \frac{8}{a\, n} \, f_n(\phi) -2 a \left( -\frac{21}{16} + \ln 2-\ln (\mu_S /m_b) \right) \,[\varrho_{n,1}]_c^{\rm LL}
\,\bigg]
\,,
\label{rhoVk2}
\end{align}
with the function $f_n(\phi)$ defined as
\begin{align}
f_n(\phi)  = &
\;\phi^2\,
 \bigg\{  
        -1+\frac{\gamma_{\rm E}}{2} + \ln (2\sqrt{n}) 
        + 2\,\sqrt{\pi}\,\phi\,\left( \frac{\gamma_{\rm E}}{2} + \ln \sqrt{n} \right) 
        - 2\,\sqrt{\pi}\,\sum\limits_{p=3}^{\infty}\,\phi^{p-1}\,\frac{\zeta_p}{\Gamma(\frac{p-2}{2})}
\nn\\[1.5 ex] & 
        -4\,\sqrt{\pi}\,\sum\limits_{p=2}^{\infty}\,\phi^p\,\frac{\zeta_p}{\Gamma(\frac{p-1}{2})}
        \,\left( \frac{1}{2}\,\Psi\left( \frac{p-1}{2} \right) -\ln \sqrt{n} \right)
+2\,\sqrt{\pi}\,\sum\limits_{p,q=2}^{\infty}\,\phi^{p+q-1}\,\frac{\zeta_p\,\zeta_q}{\Gamma(\frac{p+q-2}{2})}
\bigg\}
\,.
\label{fn}
\end{align}
We note that for the NNLL order correction terms in the moments it is sufficient
to use $a\simeq 
a^{\rm LL}=C_F\,\alpha_S^{\rm LL}$ in Eqs.~(\ref{rhokin})-(\ref{rhoVk2}). 

\end{section}

\begin{section}{Current Wilson Coefficients}
\label{Appc1}
The NLL anomalous dimension of the ${}^3S_1$ current coefficient $c_1(\nu)$ is
given by~\cite{Manohar:2000kr,Pineda:2001et,Hoang:2002yy}  
\begin{align}
 \left(\nu \frac{\partial}{\partial\nu} \ln[c_1(\nu)] \right)^{\rm NLL} 
\!\!\!\! =  
 -\:\frac{{\cal V}_c^{(s)}(\nu)
  }{ 16\pi^2} \bigg[ \frac{ {\cal V}_c^{(s)}(\nu) }{4 }
  +{\cal V}_2^{(s)}(\nu)+{\cal V}_r^{(s)}(\nu)
   + {\bf S}^2\: {\cal V}_s^{(s)}(\nu)  \bigg] 
 +\frac12 {\cal V}_{k,\rm eff}^{(s)}(\nu)
\label{c1anomdim}
\end{align}
with ${\bf S}^2=2$ for the heavy quark pair produced in the spin triplet state. The
solution at NLL order, $\xi^{\rm NLL}$, 
has been given in Eq.~(57) of Ref.~\cite{Hoang:2002yy}.
The form of NNLL ultrasoft mixing corrections in Eq.~\eqref{XiNNLLmixusoft} below 
implies the convention that the $z$ and
$\omega=1/(2-z)$ parameters in the expression for $\xi^{\rm NLL}$, as given in
Ref.~\cite{Hoang:2002yy}, are defined according to Eq.~\eqref{zLL} below.

The ultrasoft contribution to the NNLL mixing correction $\xi^{\rm NNLL}_{\rm
  m}$ in Eq.~\eqref{c1solution} 
was determined in Ref.~\cite{Hoang:2011gy} from Eq.~\eqref{c1anomdim} through
the subleading NLL ultrasoft RG evolution of the potentials ${\cal V}_2$,${\cal
  V}_r$ and ${\cal V}_{k,\rm eff}$~\cite{Hoang:2006ht,Hoang:2011gy} (see also
  Ref.~\cite{Pineda:2011aw}) and reads 
\begin{align}
\xi^{\rm NNLL}_{\rm m, usoft} & = 
\frac{2\pi \beta_1}{\beta_0^3}\,\tilde A\,\ah^2\,
 \bigg[ -\frac{7}{4}+\frac{\pi^2}{6}+z\left(1-\ln\frac{z}{2-z}\right)
      +z^2\left(\frac{3}{4}-\frac{1}{2}\ln z\right)
\nn\\[1.5 ex] & \hspace{16 ex} 
      -\ln^2\left(\frac{z}{2}\right)+\ln^2\left(\frac{z}{2-z}\right)
      -2\mbox{Li}_2\left(\frac{z}{2}\right)\bigg]
\nn\\[1.5 ex] & 
+\,\frac{8\pi^2}{\beta_0^2}\,\tilde B\,\ah^2\,
  \Big[ 3-2z-z^2-4\ln(2-z) \Big]
\,,
\label{XiNNLLmixusoft}
\end{align}
with
\begin{align}
\beta_0 &= \frac{11}{3}C_A - \frac43 T n_f \,,\quad \beta_1 = \frac{34}{3}C_A^2 - 4 C_F T n_f - \frac{20}{3}C_A T n_f\,,\\
\tilde A &= -C_F(C_A+C_F)(C_A+2C_F) \frac1{3\pi}
 \,, \label{ASchlange} \\
\tilde B &= -C_F(C_A+C_F)(C_A+2C_F) \frac{C_A(47 + 6\pi^2) -10 n_f T }{108 \pi ^2} \,, \label{BSchlange} 
\\
z & \equiv  \frac{\as^{\rm LL}}{\ah}
  \, = \,
  \bigg(1+\frac{\ah\beta_0}{2\pi}\ln\nu\bigg)^{-1}\,. 
\label{zLL}
\end{align}
%
The complete result for the soft NNLL mixing correction, $\xi_{\rm m,
  soft}^{\rm NNLL}$ is currently unknown. 
However, its linear logarithmic contribution $\propto\alpha_h^3\ln\nu$ 
was determined in Ref.~\cite{Hoang:2003ns} from the known
subleading matching conditions for the
potentials~\cite{Manohar:2000hj,Kniehl:2001ju} that appear in
Eq.~\eqref{c1anomdim} and reads 
\begin{align}
\xi^{\rm NNLL}_{\rm m, soft1} & = \frac{\ah^3}{48 \pi}\, C_F^2\, 
\bigg[\, C_A \,\Big(16\,{\bf S}^2-3\Big) 
     + 4\, C_F\,\Big(5 - 2\,{\bf S}^2\Big) - \frac{16}{5}\, T_F \,\bigg] \ln \nu\,.
\label{softmixlog}
\end{align}
The NNLL non-mixing contributions $\xi^{\rm NNLL}_{\rm nm}$ were determined in
Ref.~\cite{Hoang:2003ns} for the assignment $\mu_U=\mu_S^2/m_b$, i.e.\ for
$h=1$. For the more general case 
$\mu_U=\mu_S^2/(h m_b)$ we use in our work (see the discussion in
Sec.~\ref{sectionnewtheory}), the NNLL order non-mixing anomalous dimension
receives an additional additive term $\propto\ln h$ which reads
\begin{align}
\delta \xi^{\rm NNLL}_{\rm nm}(h) & = 
\frac{4 C_F \left(C_A^2+3 C_A C_F +2 C_F^2\right)}{3 \beta_0}\, \ah^2
\,\ln(h)\, \big(2 \ln \Big[\frac{1}{2-z}\Big]-z+1 \big) 
\,.
\end{align}
For a comparison carried out in our numerical analysis we also give the linear
soft non-mixing logarithm contained in $\xi^{\rm
  NNLL}_{\rm nm}$ for $h$=1, which is determined by expanding the full result
for $\xi^{\rm NNLL}_{\rm nm, soft}$ in $\alpha_h$: 
\begin{align}
 \xi^{\rm NNLL}_{\rm nm, soft1} & = \frac{\ah^3}{288 \pi} C_F  \bigg[
303 C_A^2+ [8 C_F (\bmS^2\!-\!12)\!-\!111 C_A]\beta _0 +2 C_A C_F (141\!-\!40 \bmS^2)\!+\!360 C_F^2
\bigg]\! \ln \nu\,.
\label{softnonmixlog}
\end{align}
Taking $\xi^{\rm NNLL}_{\rm m,soft1}$ as an approximation for the complete soft
NNLL mixing corrections, 
we can thus write the NNLL contribution to $\ln[ c_1(\nu)/c_1(1)]$
for arbitrary $h$ as 
\begin{align}
\xi^{\rm NNLL} & =\, [\xi^{\rm NNLL}_{\rm nm}(h=1) + \delta \xi^{\rm NNLL}_{\rm
  nm}(h)] + [\xi^{\rm NNLL}_{\rm m, usoft} \,+\, \xi^{\rm NNLL}_{\rm m, soft1}
]\,.
\end{align}

The matching condition for the current Wilson coefficient $c_1(1)$ is given in
Ref.~\cite{Hoang:2003ns} for $h=1$. For arbitrary $h$ we have to add a $\ln(h)$
term which arises from the term $\ln(\mu_h/m_b)$ and find 
\begin{align} \label{c1match}
c_1(1)  =&\; 1- \frac{2 C_F}{\pi}\ah +   
  \ah^2 \bigg[C_F^2\bigg(\frac{\ln 2}{3}-\frac{31}{24}
  -\frac{2}{\pi^2}\bigg) + C_A C_F\bigg(\frac{\ln 2}{2}-\frac{5}{8}\bigg) 
  + \frac{\kappa}{2} \bigg] \nn\\
&-\ah^2 C_F \ln(h)  \big[\frac{\beta _0}{\pi^2} + \frac{C_A}{2} + \frac{C_F}{3} \big] 
\,,
\end{align}
where the constant $\kappa$ was defined in Eq.~(21) of Ref.~\cite{Hoang:1998xf}.\\

The coefficient of the $v^2$ suppressed ${}^3S_1$ current $c_2(\nu)$ is needed
at the LL level and reads~\cite{Hoang:2001mm} 
\begin{align}
c_2(\nu) = -\frac{8 C_F}{3 \beta_0} \ln (\frac{\au}{\ah}) - \frac{1}{6}\,.
\label{c2}
\end{align}

\end{section}

\begin{section}{Experimental moments}
\label{AppPexp}

The experimental moments we use in this work together with their statistical and
systematic errors, respectively, read
\begin{align}
P^{exp}_{4} &=(2.17025 \pm 0.0128186 \pm 0.0464299) \times 10^{-9} \,, \nn\\
P^{exp}_{5} &=( 2.11692 \pm 0.0135132 \pm 0.0363657 ) \times 10^{-11} \,, \nn\\
P^{exp}_{6} &=( 2.13921 \pm 0.0144643 \pm 0.0317105 ) \times 10^{-13} \,, \nn\\
P^{exp}_{7} &=( 2.21297 \pm 0.0156671 \pm 0.0297788 ) \times 10^{-15} \,,\nn\\
P^{exp}_{8} &=( 2.32718 \pm 0.017121 \pm 0.029311 ) \times 10^{-17}\,,\nn\\
P^{exp}_{9} &=( 2.47683 \pm 0.0188304 \pm 0.0297155 ) \times10^{-19}\,,\\
P^{exp}_{10} &=( 2.66003 \pm 0.0208057 \pm 0.0307192 ) \times10^{-21}\,,\nn\\
P^{exp}_{11} &=( 2.87678 \pm 0.0230625 \pm 0.032201 ) \times10^{-23}\,,\nn\\
P^{exp}_{12} &=( 3.12831 \pm 0.0256223 \pm 0.0341133 ) \times10^{-25}\,,\nn\\
P^{exp}_{13} &=( 3.41679 \pm 0.0285116 \pm 0.0364466 ) \times10^{-27}\,,\nn\\
P^{exp}_{14} &=( 3.74516 \pm 0.0317625 \pm 0.0392118 ) \times10^{-29}\,.\nn
\end{align}
For our correlated fits we have determined the covariance matrix
according to the rules provided in statistics review of
Ref.~\cite{Beringer:1900zz} treating the $10$\% uncertainty in the
contribution of the continuum region for $\sqrt{s}>11.21$~GeV as a single
independent error source. Explicit formulae for the covariance matrix can also
be found in Ref.~\cite{Hoang:1998uv}.
\end{section}


%

\bibliographystyle{kp}
\addcontentsline{toc}{chapter}{Bibliography}
\bibliography{MeiBmassBib}

\end{document}